 \documentclass[article, amsmath,amssymb,floatfix, twocolumn, nopacs]{revtex4}
\usepackage{graphicx,dcolumn, bm, color, booktabs, cancel, ulem}
\DeclareMathAlphabet{\mathcal}{OMS}{cmsy}{m}{n} 
\SetMathAlphabet{\mathcal}{bold}{OMS}{cmsy}{b}{n}

\newcommand{\vomega}{\mbox{\boldmath $\omega$}} 
\newcommand{\BV}{{Brunt-V\"ais\"al\"a frequency}}        
\newcommand{\be}{\begin{equation}}  \newcommand{\ee}{\end{equation}}

\begin{document}   \title{Linking dissipation, anisotropy and intermittency in rotating stratified turbulence \\
{at the threshold of linear shear instabilities}}
\author{A.~Pouquet$^{1,2}$, D. Rosenberg$^3$, and R. Marino$^4$}
\affiliation{
$^{1}$Laboratory for Atmospheric and Space Physics, University of Colorado, Boulder, CO 80309, USA. \\
$^{2}$National Center for Atmospheric Research, P.O.~Box 3000, Boulder, CO 80307, USA.\\
$^3$ 1401 Bradley Dr. Boulder, CO 80305. \\
$^{4}$Laboratoire de M\'ecanique des Fluides et d'Acoustique, CNRS, \'Ecole Centrale de Lyon, Universit\'e Claude Bernard Lyon~1, INSA de Lyon, \'Ecully, F-69134, France. }

\begin{abstract} 
Analyzing a large data base of high-resolution three-dimensional direct numerical simulations of decaying rotating stratified flows,  we show that anomalous mixing and dissipation, marked anisotropy, and strong intermittency are all observed simultaneously in an intermediate regime of parameters in which  both waves and eddies interact 
nonlinearly.  A critical behavior governed by the stratification occurs at Richardson numbers of order unity, 
{and with the flow close to being in a state of}
   instability.
This confirms the central dynamical role, in 
{rotating stratified turbulence, of}
 large-scale intermittency, {which occurs} in the vertical velocity and temperature fluctuations, 
  as an adjustment mechanism of the energy transfer in the presence of strong waves. \\
\end{abstract}   \pacs{}              \maketitle  

 \section{Introduction, equations and diagnostics}
 
 {A signature of fully developed turbulence (FDT) is its intermittency, {\it i.e.} the occurrence of intense and sparse small-scale structures such as vortex sheets, filaments and fronts. This translates into non-Gaussian Probability Distribution Functions (PDFs) of velocity and temperature gradients, as well as chemical tracer gradients. On the other hand, the }
 atmosphere and the ocean are 
 {also}
  known for their intermittency in the  large scales, with strong wings in the  PDFs of the 
  {vertical}
  velocity and temperature fields
 {themselves. Such high vertical velocities}
 are observed in the nocturnal Planetary Boundary Layer\cite{lenschow_12}; 
 {this leads to}
 strong spatial and temporal variations of the rate of kinetic energy dissipation, as 
 {measured}
in oceanic ridges\cite{klymak_08, vanHaren_16j}.
{Similarly, micro-structures, observed in the frontal Antarctic Circumpolar Current are formed by quasi-geostrophic eddies flowing over bottom topography\cite{zheng_19}.
These structures are due to the bathymetry which has recently been assessed with increased accuracy \cite{kar_18}. Such interactions between turbulence and stratification can affect many processes in the atmosphere and the ocean, such as for example rain formation \cite{shaw_01}, or the lifetime of melting ice 
shelves\cite{vreugdenhil_19}.}

Large-scale intermittency is  also found in  high-resolution Direct Numerical Simulations (DNS) of stratified flows, 
{with or without}
 rotation\cite{rorai_14, feraco_18}, with a direct correlation to high levels of dissipation.
 However, isotropy is classically assumed when estimating energy dissipation of turbulent flows, from laboratory experiments to oceanic measurements, 
 {even though}
  it has been known for a long time that small-scale isotropy recovers slowly in terms of the controlling parameter, such as in wakes, boundary layers, and pipe or shear flows. 

A lack of isotropy can be associated with intermittency, 
{as well as }
 with the long-range interactions between large-scale coherent structures and small-scale dissipative 
 eddies\cite{browne_87}. 
In the purely rotating case, vertical Taylor columns form and, using particle image velocimetry, space-time dependent anisotropy has been shown to be important\citep{campagne_15}.
{Spectra follow weak turbulence arguments for strong rotation\cite{sharma_18}, and pressure acts on nonlocal interactions between strong vortices at small scales and large-scale fluctuations \cite{yang_18}.}
In the case of pure stratification, its  role  on small-scale anisotropy was studied experimentally in detail\cite{thoroddsen_92}. 
Spectral data and dissipation data are mostly stream-wise anisotropic because of the shear, on top of the anisotropy induced by the vertical direction of stratification\cite{werne_01}. 
The vertical integral length scale does not grow, contrary to its horizontal counterpart\cite{godeferd_03}, and vertical scales are strongly intermittent. 

Different components of the energy dissipation tensor have been evaluated, 
for purely stably stratified flows
{or wall turbulence,}
 as a function of governing parameters\cite{smyth_00b, gerolymos_19, garanaik_19, lang_19}, and a slow return to isotropy is found only for rather high buoyancy Reynolds number
{\cite{smyth_00b},}
 of the order of $R_B\approx 10^3$  (see next section for definitions of parameters). 
With strong imposed shear and using anisotropic boxes, anisotropy is found to be strongest when turbulence is weakest, as expected, and anisotropic eddies in the small scales depend on the effective scale-separation 
{between the large-scale containing eddies, or the buoyancy scale for stratified flows, and the dissipative scale of the turbulence\cite{smyth_00b}.}
Part of the difficulty in assessing the return to isotropy in either the large or the small scales, however, is that there is a strong coupling between scales, through the interactions of gravity waves and fine-structure shear 
layers\cite{fritts_09b}, as well as in fronts. 

In this context,  we evaluate quantitatively the link between mixing and dissipation, anisotropy and intermittency in the presence of both rotation and stratification, and as a function of the intensity of the turbulence. This is accomplished in the framework of  a large series of unforced DNS runs for the Boussinesq equations. 
With ${\cal P}$ the total pressure, ${\bf u}={\bf u}_\perp+w\hat e_z$ the velocity, $\theta$ the temperature fluctuation (normalized to have dimensions of a velocity), and $\nabla \cdot {\bf u} =0$ because of incompressibility, we have in the unforced case:
 \begin{eqnarray} 
 \frac{\partial {\bf u}}{\partial t} + \mbox{\boldmath $\omega$} \times {\bf u} + 2 \mbox{\boldmath $\Omega$} \times {\bf u}  &=& -N \theta \hat e_z  - \nabla {\cal P} + \nu \nabla^2 {\bf u}, \\ 
  \frac{\partial\theta}{\partial t} + {\bf u} \cdot \nabla \theta &=&  Nw + \kappa \nabla^2 \theta  \ ;\label{eq:momentum} 
  \end{eqnarray}
 $\nu$ is the viscosity, $\kappa$ the diffusivity, $\vomega=\nabla \times {\bf u}$  the vorticity and $N$ the \BV.
Rotation, of intensity $\Omega=f/2$, and stratification are in the  vertical ($z$) direction.

We use the pseudo-spectral  Geophysical High Order Suite for Turbulence (GHOST) 
code with hybrid MPI/OpenMP/CUDA parallelization and linear  scaling up to at least 130,000 
cores\cite{hybrid_11}. The GHOST-generated database considered here consists of fifty-six simulations on grids of $1024^3$ points, as well as three at $512^3$, twelve at $256^3$, and two at $128^3$ resolutions, all in a triply periodic box\cite{rosenberg_15, pouquet_18}.
 Initial conditions for most runs are isotropic in the velocity; thus at t=0, $w/u_\perp \lesssim 1$, 
  and we take zero temperature fluctuations, so that $\theta$ develops in a dynamically consistent way. Initial conditions in quasi-geostrophic (QG) equilibrium have also been  considered\cite{rosenberg_15}, in that case with $N/f\approx 5$, $w(t=0)=0$ and $\theta(t=0)\not= 0$.
 The analysis of the QG set of runs, indicated in the figures by star symbols, has not introduced any major change in the conclusions\cite{pouquet_18}, although it displays more intermittency and anisotropy (see Figs. \ref{f:4} and  \ref{f:5} below). Finally, with $\perp$ referring to  the horizontal direction, $k=\sqrt{|{\bf k}_\perp|^2 + k_z^2}$ is the isotropic wavenumber. 

The dimensionless parameters of the problem are the Reynolds, Froude,  Rossby and Prandtl numbers: 
 \be
Re=\frac{U_0L_{int}}{\nu}\ , \ \ Fr=\frac{U_0}{L_{int}N}\ , \ \ Ro=\frac{U_0}{L_{int}f}\ , \ \ Pr=\frac{\nu}{\kappa},
 \label{PARAM}     \ee
where $U_0$ is the {\it rms} velocity and 
{$L_{int}=\int [E_v(k)/k]dk/E_v$ is the isotropic}
  integral scale, both evaluated at the peak of dissipation. For all runs,  we set $Pr=1$. The kinetic, potential and total energies $E_v, E_p$ and $E_T=E_v+E_p$, of respective isotropic Fourier spectra $E_{v,p,T}(k)$, and their dissipation rates ${\bar{\epsilon}}_{v,p,T}$ are: 
\be 
E_v= \left<|{\bf u}|^2/2\right>, {\bar{\epsilon}}_v= DE_v/Dt=\nu {Z_V, \ Z_V=\left<|\vomega|^2\right>,} \ee
\be
E_p= \left< \theta^2/2\right>, {\bar{\epsilon}}_p=DE_p/Dt=\kappa {Z_P, \ Z_P=\left<|\nabla \theta|^2\right>,}
\label{E}  \ee 
with ${\bar{\epsilon}}_T={\bar{\epsilon}}_v+{\bar{\epsilon}}_p$,
{and where $Z_{V,P}$ are the kinetic and potential enstrophies.}
 Spectra can also be expressed in terms of $k_\perp$ or $k_z$ (as in equation (\ref{tensors}) below).
{It may also be useful to define other derived parameters. For example, the}
  Richardson number $Ri$, buoyancy  Reynolds number $R_B$, buoyancy interaction parameter $R_{IB}$ and gradient Richardson number $Ri_g$ are written as:
 \begin{eqnarray} 
Ri&=& (N/S)^2 \ \ , \  \  R_B=ReFr^2 ,  \\
R_{IB}&=& {\bar{\epsilon}}_v/(\nu N^2)\ , \ Ri_g= N(N-\partial_z\theta)/S({\bf x})^2,   
 \label{RB}   \end{eqnarray} 
 with $S=\left<\partial_z u_\perp \right>$ representing  the internal shear that develops in a dynamically consistent way,
 {and in the absence of imposed external shear.}
 $Ri_g$ is a point-wise measure of instability; it can be negative when the vertical temperature gradient is locally larger than the (constant) \BV, indicative of strong local overturning. 
{We should note here that different definitions can be found in the literature. In particular, the buoyancy Reynolds number is often expressed as\cite{ivey_08} ${\bar{\epsilon}}_v/[\nu N^2]$, corresponding to $R_{IB}$ here. The distinction between $R_B$ and $R_{IB}$ is physically important. Indeed, we can}
 also define $\beta$ as a global measure of the efficiency of kinetic energy dissipation, with respect to its dimensional evaluation $\epsilon_D {\sim U_0^3/L_{int}}$:
 \begin{eqnarray}
\beta&\equiv&{\bar{\epsilon}}_v/\epsilon_D= \tau_{NL}/T_v, \\
\tau_{NL}&=&L_{int}/U_0, \ T_v=E_v/{\bar{\epsilon}}_v\ .
\label{eq:beta}  \end{eqnarray}  
$\tau_{NL}$ and $T_v$  
{are}
the two characteristic times defining nonlinear transfer and energy dissipation; one can also define  the waves periods as $\tau_{BV}=2\pi/N$ and $\tau_f=2\pi/f$. 
{It follows that, in the intermediate regime of wave-eddy interactions, one has the following simple relationship :
\be R_{IB}=\beta R_B. \ee
This can be  justified through  a dimensional argument corroborated by numerical results\cite{pouquet_18}.}
Note that, in  fully developed turbulence, one has $T_v=\tau_{NL}$
{and $\beta=1$.}
 We also showed  that the characteristic times associated with the velocity and temperature and based on their respective dissipation rates, $T_v$ and $T_p=E_p/{\bar{\epsilon}}_p$, vary substantially with  governing parameters, being comparable  in a narrow range of Froude numbers when large-scale shear layers destabilize\cite{rosenberg_16}.
The direct numerical simulations cover a wide range of parameters\cite{marino_15w, rosenberg_16, pouquet_18}: 
\begin{equation*}
10^{-3} \le Fr \le 5.5 \ , \  2.4\le N/f \le 312 \ , \ 1600\le Re \le 18590. \end{equation*}
$R_B$ and $R_{IB}$ vary roughly from  $10^{-2}$ to $10^5$, values  which, at the upper end, are relevant to the ocean and atmosphere. A few purely stratified runs are considered as well. 
 
\begin{figure*} 
 \includegraphics[width=7.5cm,height=6cm]{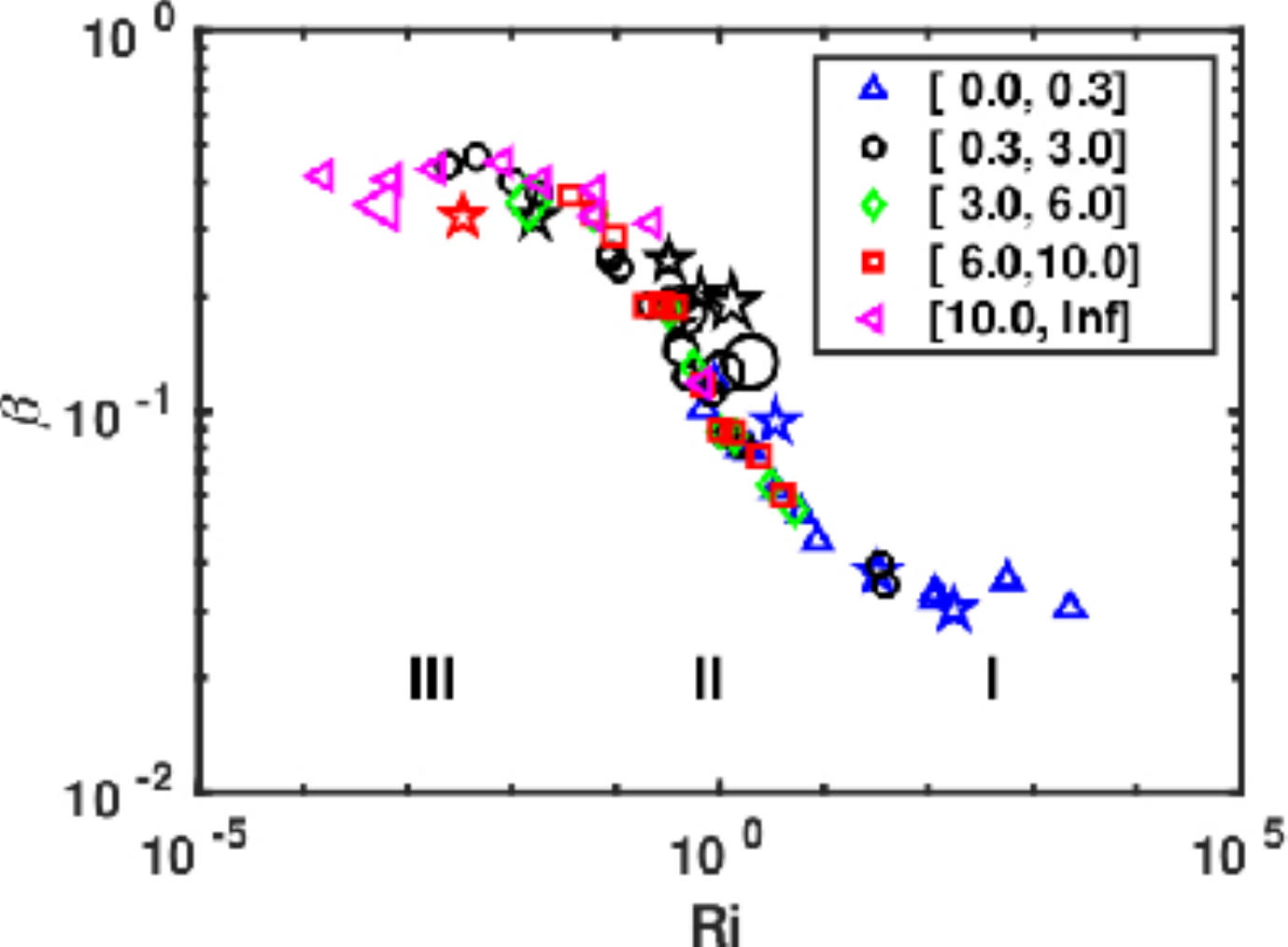}  \hskip0.15truein 
 \includegraphics[width=7.5cm,height=6.0cm]{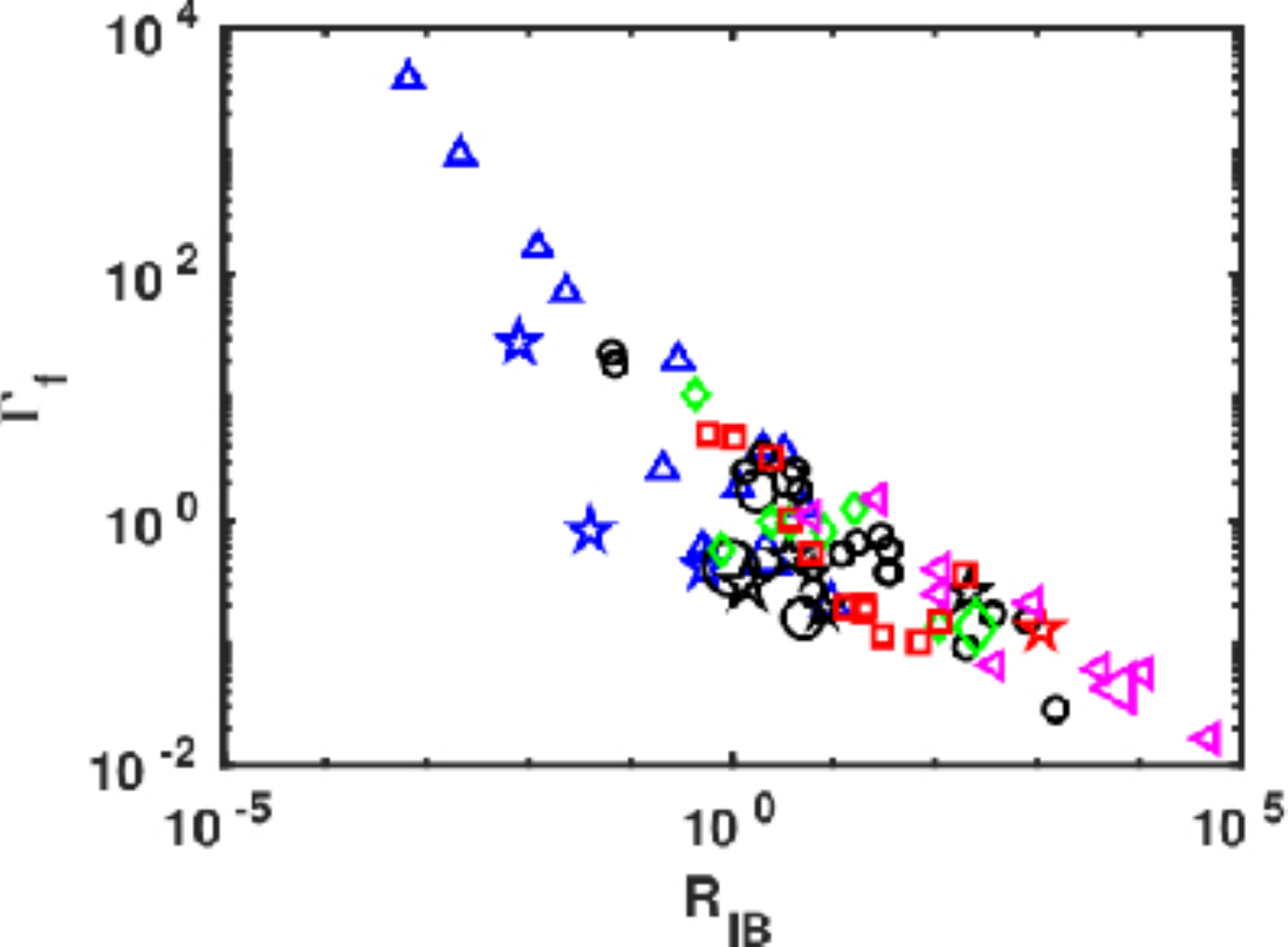}
\caption{ 
Left:  Variation with Richardson number of the kinetic energy dissipation  efficiency $\beta$. The Roman numerals at the bottom delineate the three regimes of rotating stratified turbulence\cite{pouquet_18}.
Right: Variation with buoyancy interaction parameter 
{$R_{IB}$}
 of the mixing efficiency $\Gamma_f$
 {defined in equ. (\ref{Gamma}).} 
Colored symbols indicate Rossby number ranges (see inset).  
}\label{f:1}    \end{figure*}

Anisotropy has been studied extensively for a variety of flows\cite{sagaut_08a}, and many diagnostics have been devised. Here, we concentrate on the following set, 
{starting with the integral scale,  the subscript $\mu$ } representing $z,\perp$: 
\begin{equation} 
\frac{L_{int,\mu}}{2\pi}=\frac{\Sigma k^{-1}_{\mu} E_{v}(k_\mu)}{\Sigma E_{v}(k_\mu)}, \label{Lint} \ee
{with $L_{int}$ representing
 the integral scale for the isotropic case, that is in terms of isotropic wavenumber.}
Integral scales 
{are}
 known to increase with time in FDT, and it has been shown to do the same in rotating and/or stratified turbulence. This is a manifestation of the interactions between widely separated scale that feed the large-scale flow through what is known as eddy noise together with, in the rotating case in the presence of forcing, the occurrence of an inverse cascade of energy. 
 
\begin{figure*} 
\includegraphics[width=5.0cm,height=5cm]{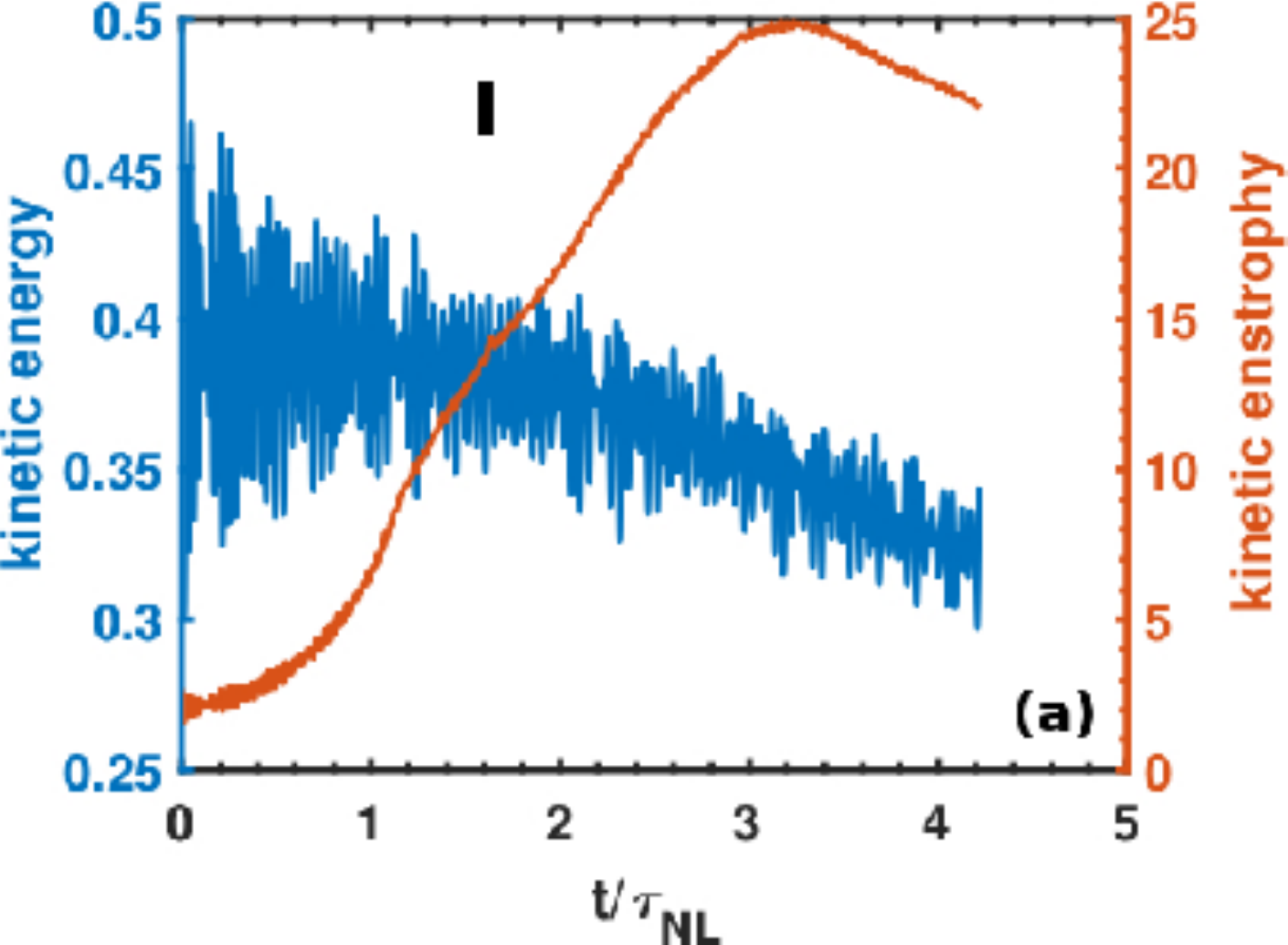}  \hskip0.15truein 
\includegraphics[width=5.0cm,height=5cm]{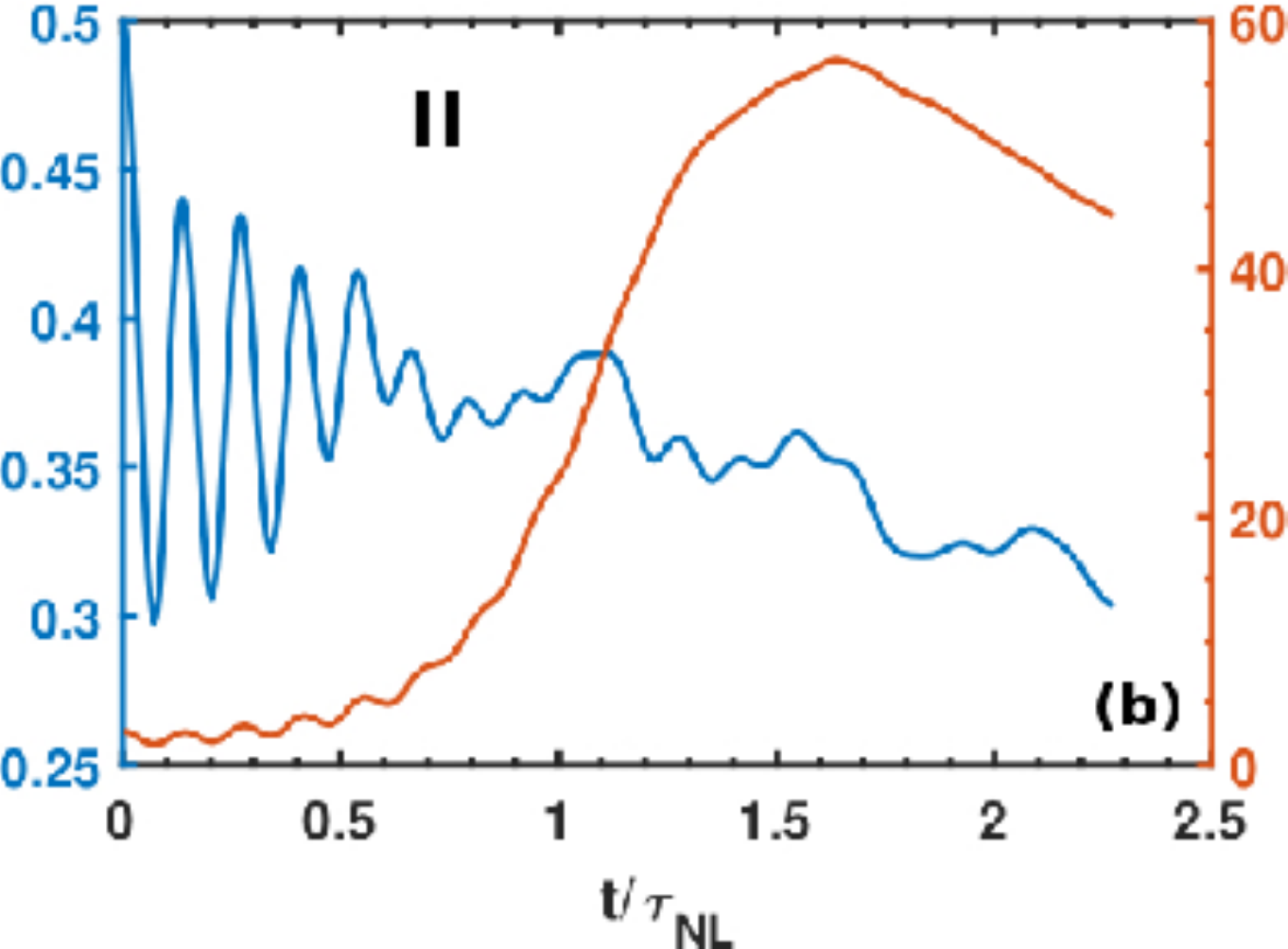}  \hskip0.15truein  
 \includegraphics[width=5.0cm,height=5cm]{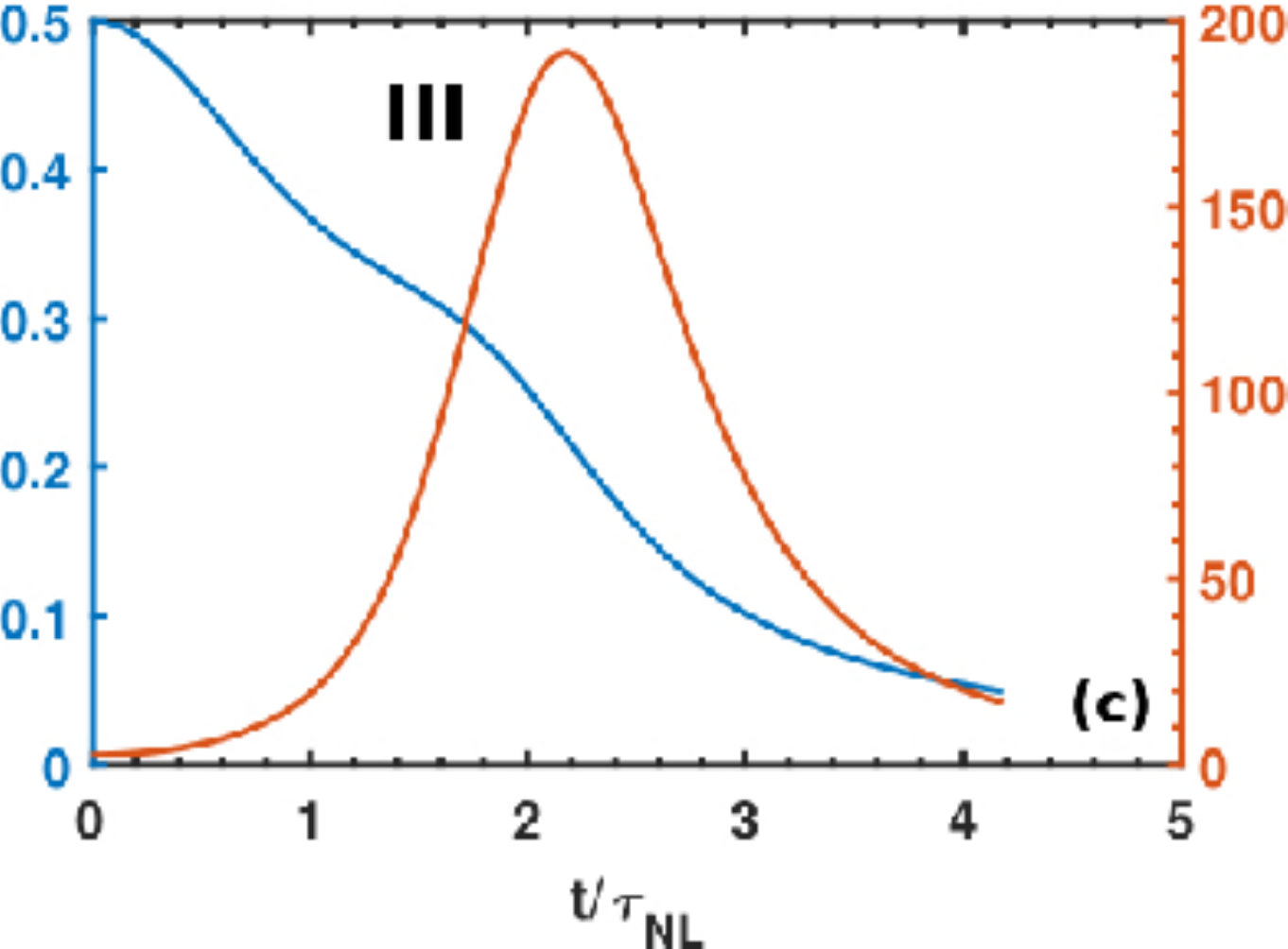}  \vskip0.05truein 
\includegraphics[width=5.0cm,height=5cm]{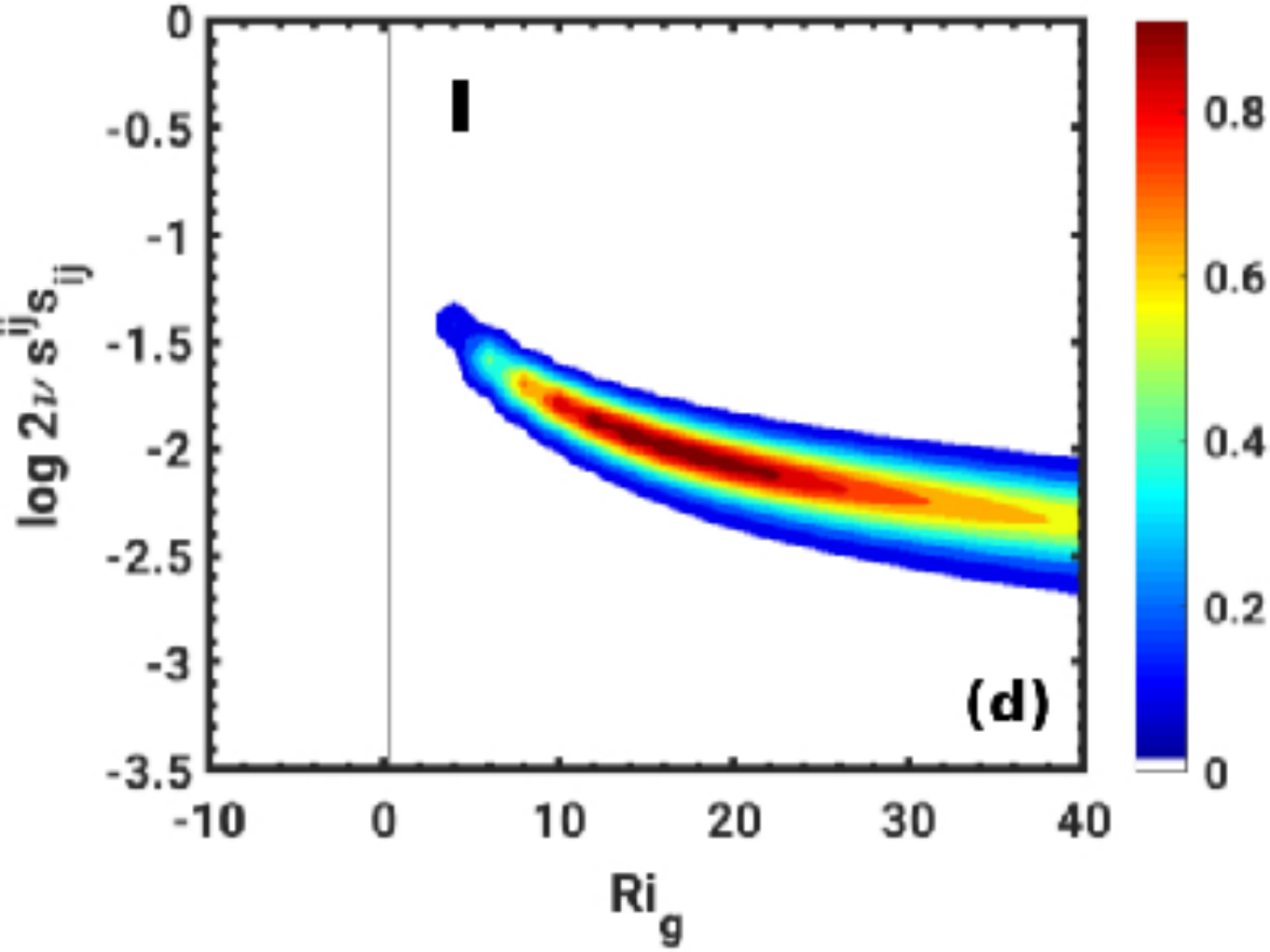}  \hskip0.15truein  
\includegraphics[width=5.0cm,height=5cm]{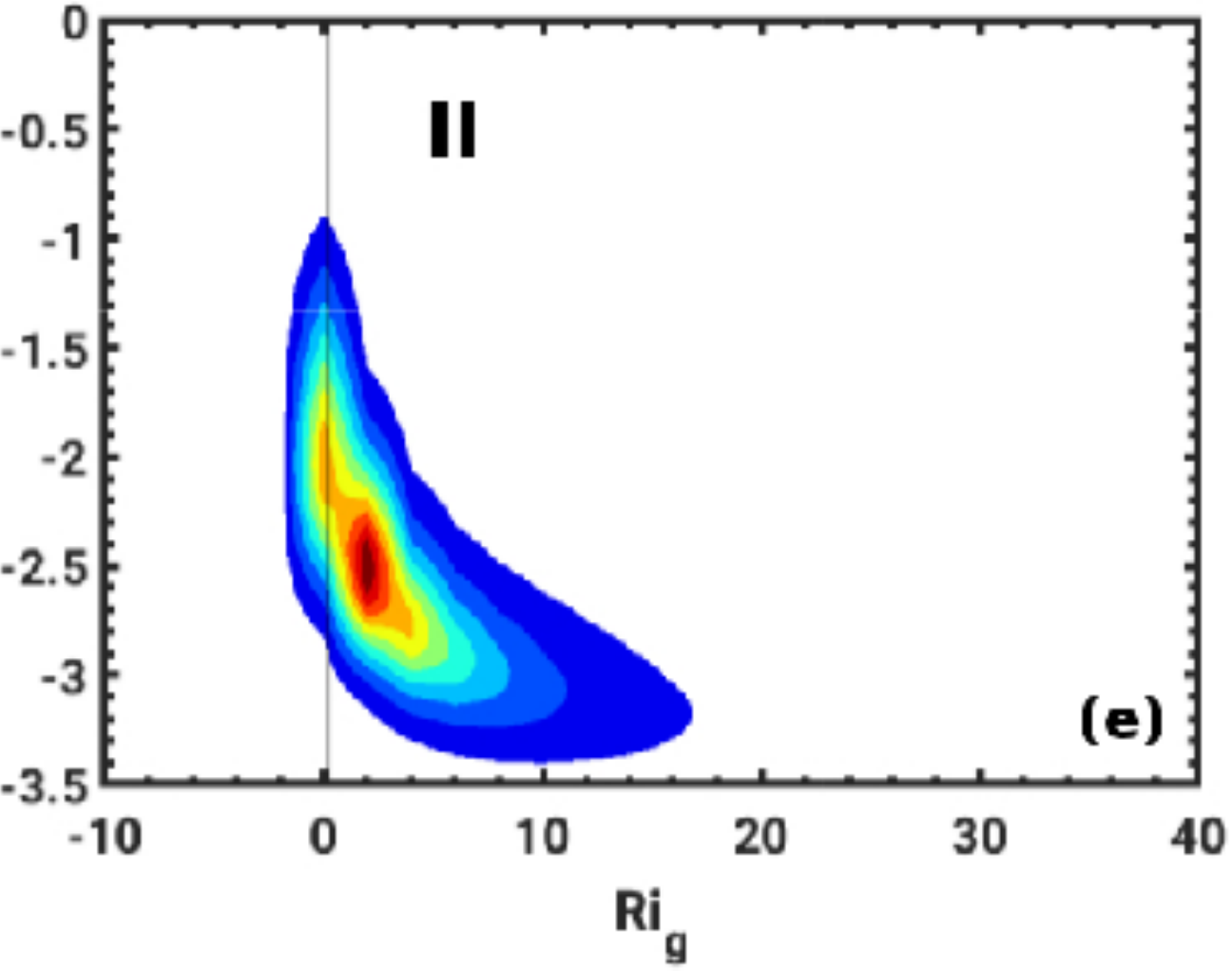}  \hskip0.15truein  
 \includegraphics[width=5.0cm,height=5cm]{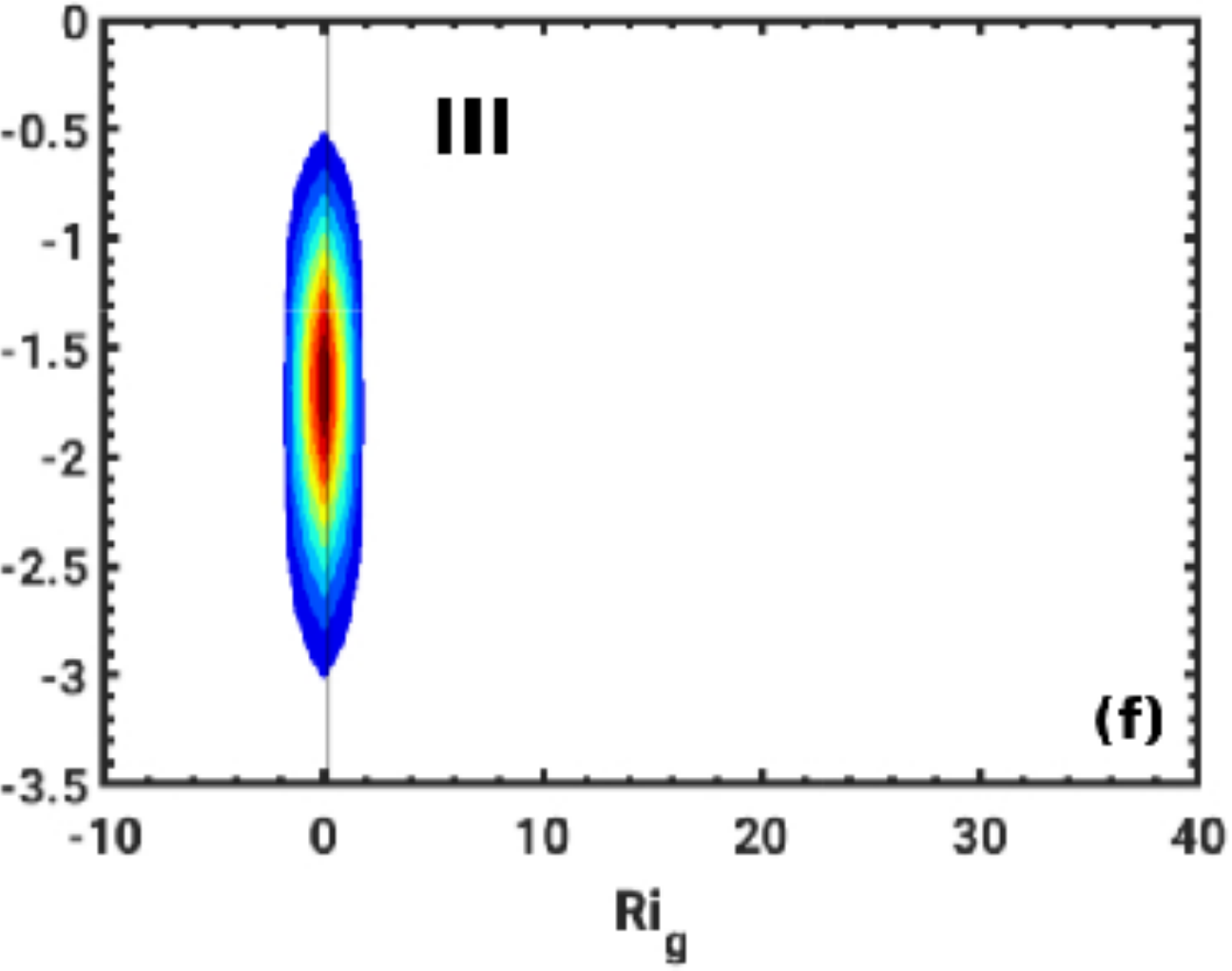}  \vskip0.15truein  
\caption{ 
{(a)-(c): Temporal variations,  in units of turn-over time,
 of kinetic energy (left, blue axis in each plot) and of kinetic enstrophy $Z_V=\left< |\vomega|^2 \right>$ (right, red axis in each plot). All runs are performed on grids of $1024^3$ points and are in one of the three regimes identified in Fig. \ref{f:1} (see text for parameters): I (a), II (b) and III (c).}
{(d)-(f):} Joint PDFs of point-wise kinetic energy dissipation $\epsilon_v({\bf x})$ and gradient Richardson number $Ri_g({\bf x})$ 
{for the same three runs as above.}
$Ri_g=1/4$
{is indicated by the thin vertical lines. All  plots use the same color bar given at left.}
}\label{f:2}    \end{figure*}

 {Other signatures of anisotropy can be obtained through the properties of the following tensors:}
\be 
b_{ij}= \frac{\left< u_iu_j \right>}{\left< u_k u_k \right>} - \frac{\delta_{ij}}{3}, \ 
d_{ij}=\frac{\left< \partial_k u_i \partial_k u_j \right>}{\left< \partial_k u_m \partial_k u_m \right>} - \frac{\delta_{ij}}{3}, \ee
\be
g_{ij}=   \frac{\left< \partial_i \theta\  \partial_j \theta \right>}{\left< \partial_k \theta\  \partial_k \theta \right>} 
         -  \frac{\delta_{ij}}{3}, \   
v_{ij}=  \frac{\left<\omega_i \omega_j \right>}{\left<\omega_k \omega_k \right>} -\frac{\delta_{ij}}{3}  \ .
 \label{tensors} \ee 
{These tensors are equal to zero in the isotropic case.}
For reference, we also write the point-wise dissipation, $\epsilon_v ({\bf x})= 2\nu s_{ij} s^{ij}$, where 
$s_{ij}({\bf x}) = \frac{1}{2}(\partial_i u_j + \partial_j u_i)$ is the strain rate tensor.  
We define as usual the second and third-order invariants of a tensor $T_{ij}$ as $T_{II}=T_{ij}T_{ji}$ and 
$T_{III}= T_{ij}T_{jk}T_{ki}$. For the tensors above, they are denoted respectively 
$b_{II,III}$, $d_{II,III}$,  $g_{II,III}$, and $v_{II,III}$ (see for example Refs. \citet{browne_87, antonia_91, smyth_00b, sagaut_08a} for details and interpretation). They refer in particular to the geometry of the fields (one-dimensional or 1D {\it vs.} 2D, 3D, and axisymmetric, oblate or prolate).
In what follows, all anisotropy tensors and their invariants are computed from a snapshot
of the data cube at the peak of total enstrophy  
{$Z_T=\left< |\vomega|^2 + |\nabla \theta|^2 \right>$}
{(and thus, at the peak of dissipation ${\bar {\epsilon_T}}$)}
for each run, as are all PDFs and
quantities associated with buoyancy flux, {\it e.g.}, $\Gamma_f$ defined in the next section. All other 
quantities that are plotted are computed based on spectra that are 
averaged in time over the peak in enstrophy. 
{Specifically, the chosen time intervals, different for different runs, are taken so that the variation of the total enstrophy in each case is no more than 2.5\%  from its peak value} when the turbulence is fully developed. This ensures a lack of correlation between data points within the parametric study. Note that most of the symbols used throughout 
the paper, together with their definitions, are provided for convenience in 
Appendix Table \ref{appendix}.

 \section{At the threshold of shear instabilities}     \label{S:threshold}
 
Rotating stratified turbulence (RST) consists of an ensemble of interacting inertia-gravity waves and (nonlinear) eddies. 
It can be classified into three regimes, I, II, and III, with dominance of waves in I for small Froude number
{(and small $R_{IB}$),}
 and dominance of eddies in III for high $R_{IB}$: then, the waves play a secondary role and dissipation recovers its fully developed turbulence isotropic limit  ${\bar{\epsilon}}_D$, within a factor of order 
 unity\cite{djenidi_17}. In the intermediate regime(regime II), one finds (i) $\beta \sim Fr$, as required by weak turbulence arguments; this  is the first central result in Ref. \citet{pouquet_18}, together with the following two  other laws: (ii) kinetic and potential energies are proportional (but not equal), with no dependence on governing parameters in regime II where waves and nonlinear eddies strongly interact; and (iii) similarly for the ratio of vertical to total kinetic energy, $E_z/E_v$.

With these three constitutive laws,  
{(i)-(iii),}
one can recover and establish a large number of scaling relationships, such as the ratio of characteristic 
scales\cite{pouquet_18}, or for the mixing efficiency defined as:
\be \Gamma_f\equiv B_f/{\bar{\epsilon}}_v,\  B_f=N\left<w\theta \right>, \label{Gamma} \ee
 $B_f$ being the buoyancy flux. One finds 
$\Gamma_f \sim R_B^{-1}\sim Fr^{-2}$ in regimes I and II, and $\sim R_{IB}^{-1/2}\sim Fr^{-1}$ in regime III. Such scalings, predicted from simple physical arguments in \cite{pouquet_18} have been observed at high $R_{IB}$, for example in oceanic data\cite{mashayek_17}.
Defining $\Gamma_\ast \equiv {\bar{\epsilon}}_p/{\bar{\epsilon}}_v$ 
{as the reduced mixing efficiency}
provides another simple measure of irreversible mixing by looking at how much dissipation occurs in the potential and kinetic energy respectively. It is easily shown using the  laws given above that, for the saturated regime III, $\Gamma_\ast \sim Fr^{-2}$ since the Ellison scale $L_{Ell}=2\pi \theta_{rms}/N$ becomes comparable to $L_{int}$ in that case (see Fig. 6 in that paper).
These scaling  laws extend smoothly to  the purely stratified flows we have analyzed, where, for regime II, the reduced mixing efficiency was found  to 
{vary}
 linearly with Froude number \cite{feraco_18}. These results are also compatible with other results obtained for that case\cite{shih_05, ivey_08, maffioli_16b, garanaik_19, lang_19}.

We thus begin our investigation by examining mixing and dissipation. We show 
in Fig. \ref{f:1} the dissipation efficiency $\beta={\bar {\epsilon}}_v/\epsilon_D$ as a function of 
Richardson number. Unless specified otherwise, data is binned in 
Rossby number (refer to the legend in Fig. \ref{f:1}(left)), as in 
most subsequent scatter plots, 
with roughly the same number of runs in each bin. For runs 
initialized with random isotropic conditions, the color {\it and} 
symbol of a given data point both indicate 
 which Rossby number bin it resides in. Star symbols 
 {are used for} 
quasi-geostrophic initial conditions, with a balance between pressure gradient, Coriolis force and gravity, and the color alone indicates the bin range it belongs to. For all scatter plots, the size of a symbol is proportional to the viscosity of the run, with the smallest symbols denoting runs on grids of $1024^3$ points and higher Reynolds numbers, and the two largest symbols denoting the two runs on grids of $128^3$ points at the lowest $Re$. 

Note in the plot of $\beta(Ri)$ the presence of an inflection point for 
$Ri\lesssim 1/4$, and the two plateaux  starting at $Ri\approx 10^{-2}$ 
and $\approx 10$ with an approximate scaling $\beta\sim Ri^{-1/2}$ in 
the intermediate regime, consistent with $\beta\sim Fr$, as found before\cite{pouquet_18}. 
As stated earlier, this defines the three regimes 
of rotating stratified turbulence, I, II and III, in a similar fashion 
as for the case of purely stratified turbulence\cite{ivey_08}.

The mixing efficiency $\Gamma_f$ is plotted in Fig. \ref{f:1} (right) as a function of buoyancy interaction parameter 
{$R_{IB}$. The three data points with $R_{IB} \gtrsim 10^4$ have Froude numbers above unity.}
It  also follows approximately two scaling laws. It can become singular in the quasi-absence of kinetic energy dissipation (when measured in terms of buoyancy flux), and indeed $\Gamma_f$ takes high values for the runs at low $Fr$. Its slower decay with $R_{IB}$ for strongly turbulent flows starts at a pivotal value of $R_{IB}\approx 1$, a threshold which will be present in most of the data analyzed herein. The decay of $\Gamma_f$ to low values is inexorable in the absence of forcing and with zero initial conditions in the temperature field which, at high $R_{IB}$, becomes decoupled from the velocity and evolves in time in a way close to that of a passive scalar.

\begin{figure*}   
\includegraphics[width=5.0cm,height=5.0cm]{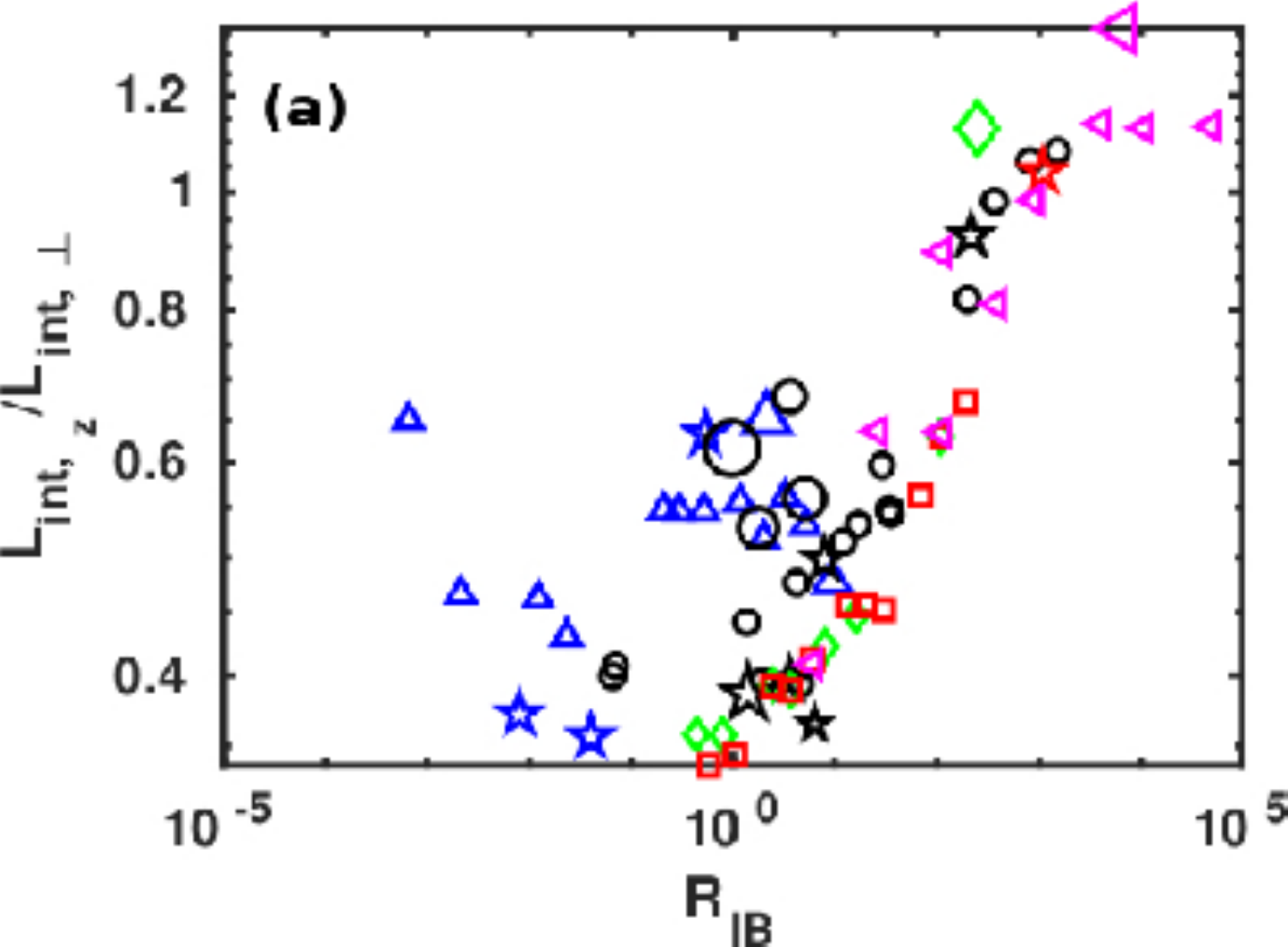}   \hskip0.1truein 
\includegraphics[width=5.0cm,height=5.0cm]{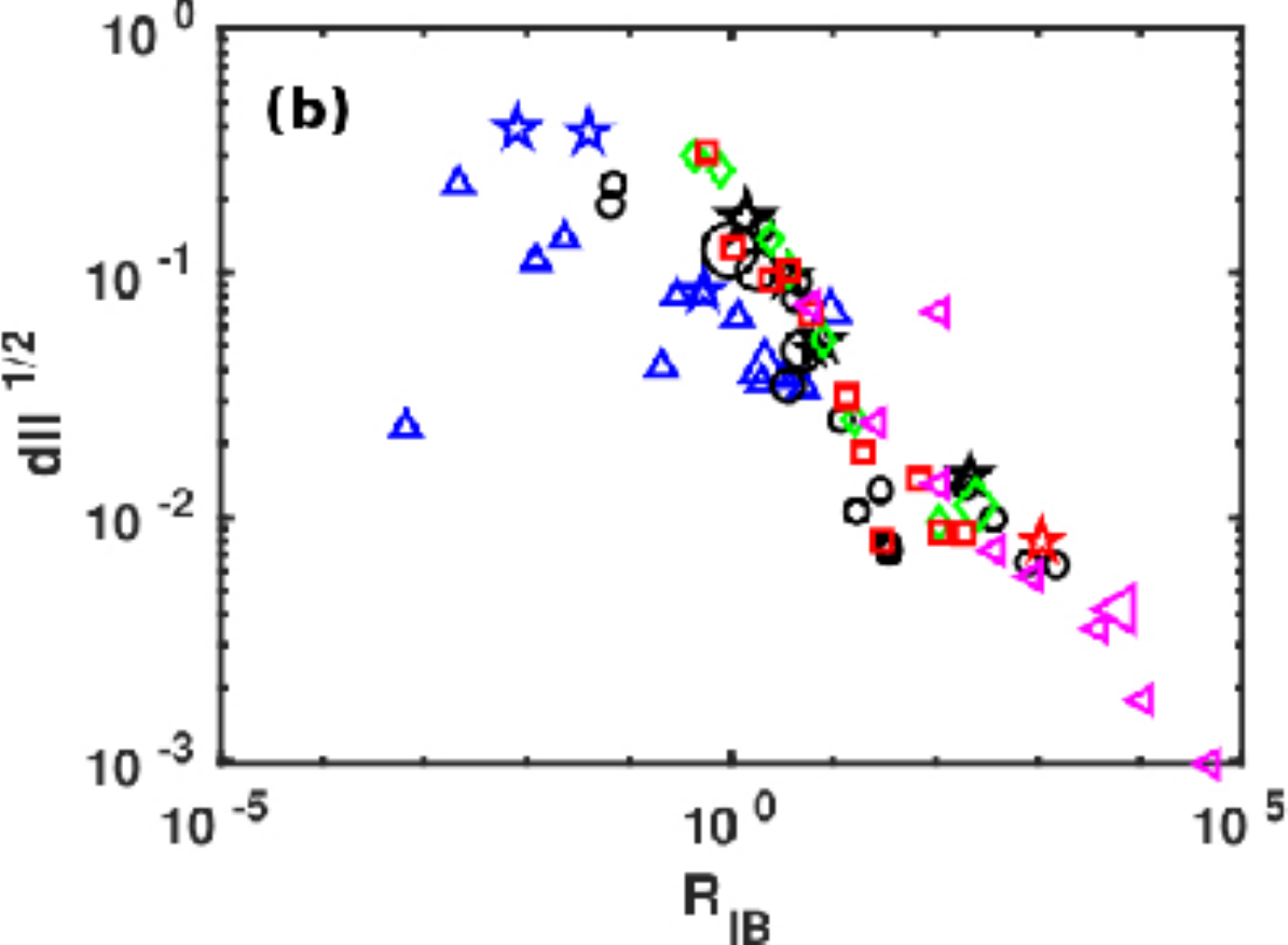}   
\includegraphics[width=5.0cm,height=5.0cm]{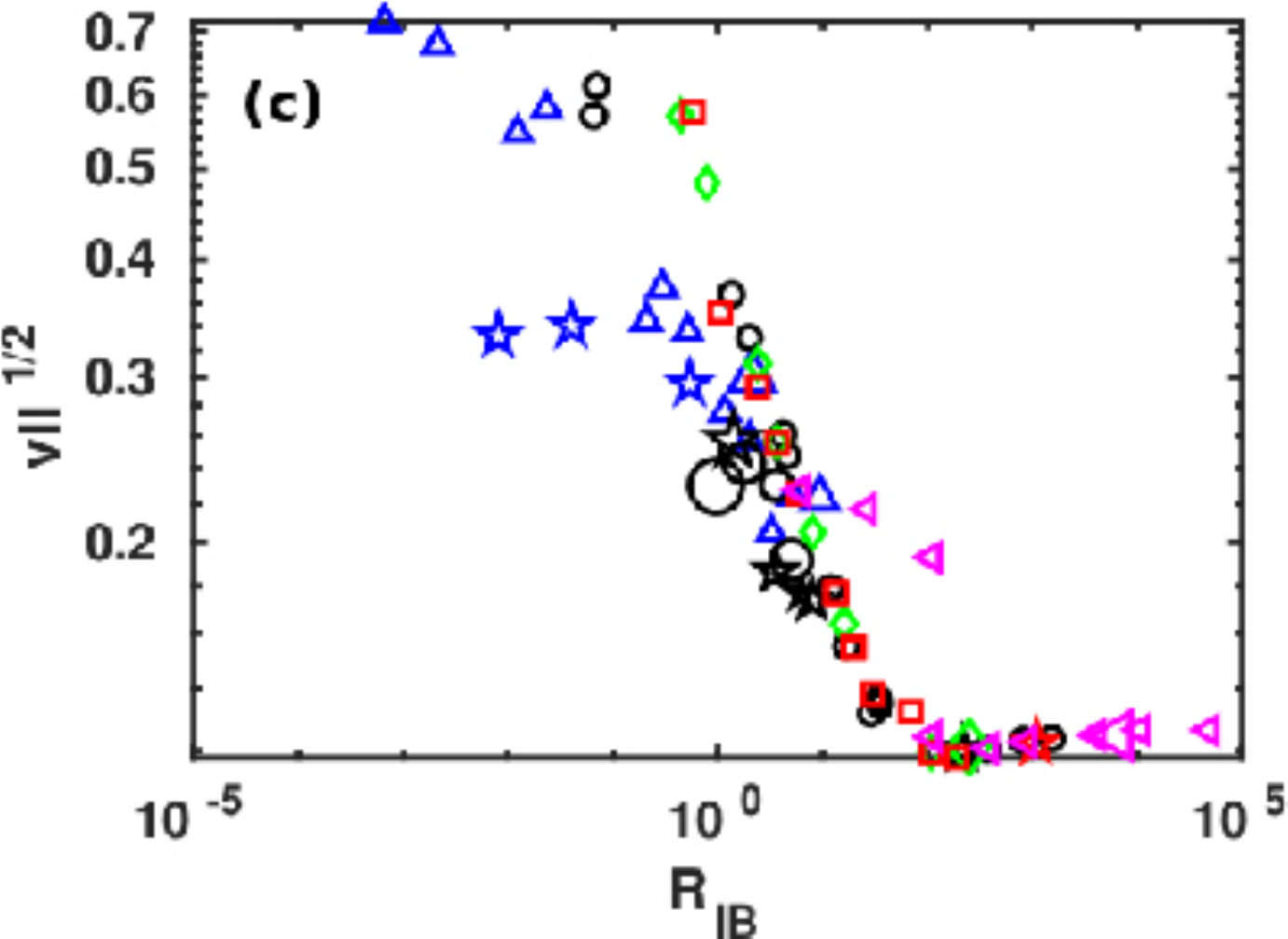}   
\caption{
As a function of buoyancy interaction parameter $R_{IB}={\bar{\epsilon}}_v/[\nu N^2]$, we plot: 
(a) Ratio of vertical to horizontal integral scales (see equation (\ref{Lint}); 
(b) $d_{II}^{1/2}$; and 
(c) $v_{II}^{1/2}$ 
(see equations (\ref{tensors}) for definitions of second tensor invariants for the velocity and vorticity,
{$d_{II}^{1/2}$ and $v_{II}^{1/2}$). Binning is in Rossby number.}
}\label{f:3}    \end{figure*}

{Figure \ref{f:2} (a)-(c) displays the variation with time of the kinetic energy and enstrophy, $E_v$ and $Z_v$, where the time is expressed in units of turn-over time, $\tau_{NL}$ defined in equation (\ref{eq:beta}). The specific runs are computed on grids of $1024^3$ points. There is roughly a factor of 10 in Froude number from regimes I to II, and from II to III; specifically, we have:
Run  5, with $Fr\approx 0.007$, $N/f\approx 31$, $Re \approx 14000$, ${\cal R}_B\approx 0.75$ in regime I; 
Run 32, with $Fr\approx 0.07$, $N/f\approx 42$, $Re \approx 12200$, ${\cal R}_B\approx 65$ in regime II; and 
Run 58, with $Fr\approx 0.89$, $N/f\approx 2.5$, $Re \approx 4700$, ${\cal R}_B\approx 3760$ in the third 
regime\cite{rosenberg_15, pouquet_18}. 
Note the different scales on both axes, and the different ranges of values for enstrophy in the three regimes: there is more enstrophy (and hence more dissipation) as we move from regime I to regime III. There are fast oscillations in the first regime (Fig. \ref{f:2} (a)). They are a signature of the fast exchanges (compared to the turn-over time) of energy between the kinetic and potential modes. 
In regime II (Fig, \ref{f:2} (b)), the oscillations are slower and become more complex once the maximum of enstrophy is reached and the flow is a superposition of nonlinearly interactive modes.
Finally, the higher enstrophy values  in the last regime (Fig. \ref{f:2} (c)) is related to strong small-scale dissipative structures. 
 The maximum of kinetic enstrophy (and thus of kinetic energy dissipation) is reached at a later time in regime I than in the other two regimes, corresponding to a slower development of small scales through weak nonlinear mode coupling. Also, as expected, the  energy decays faster as we approach the fully turbulent regime.
Similar results hold for the potential energy and its dissipation (not shown).

Joint PDFs of the point-wise gradient Richardson number and kinetic energy dissipation, for the same runs as in the top panels, are  shown in Fig. \ref{f:2} (d)-(f). The threshold of shear instability, $Ri_g=1/4$, is indicated in all three plots by a thin vertical line. For regime II (Fig. \ref{f:2} (e)), most points in the flow are close to (but still slightly above) the threshold, $Ri_g\gtrsim1$. The point-wise kinetic energy  dissipation is  centered on $\approx 3 \times 10^{-3}$ but covers locally a range of values almost  two orders of magnitude wide for $Ri_g\approx 1$. For runs in regime I (Fig. \ref{f:2} (d)), no data point reaches $Ri_g=1$, and rather dissipation values are found in a narrow band extending to high $Ri_g$. On the other hand, in the opposite case of strongly turbulent flows (regime III in Fig. \ref{f:2} (f)),
 the bulge of points around $Ri_g\approx 1/4$ is much narrower with a flow almost everywhere at the brink of linear instability,  reaching higher values of local dissipation, and with a larger extension in its local values, over 2.5 orders of magnitude; this can be seen as being indicative of intermittent behavior, as we shall analyze below in Fig. \ref{f:4}.
This high number of data points, for a given set-up, with local gradient Richardson numbers close to $1/4$ has been noted before by several authors. For example,  an equivalent result based on an earlier analysis of oceanic data\cite{moum_89} (see Figure 15) shows that in that case, most of the points are centered at $Ri_g\approx 0.4$, with also roughly four orders of magnitude  in the variation of the local energy dissipation rate. Such a high density of values for $Ri_g\approx 1/4$ has recently been interpreted as a manifestation of self-organized criticality\cite{smyth_19}, with flow destabilization occurring in a wide range of intensity displaying power-law behavior, as analyzed on observations of oceanic microstructures.}

{We  now focus on the anisotropy of these flows.}
Large-scale anisotropy can be measured by the ratio 
$L_{int,z}/L_{int,\perp}$.
 As shown in Fig. \ref{f:3}(a), it increases with $R_{IB}$ at a slow rate, starting at $R_{IB}\approx 1$  before settling sharply to a value close to unity for  high $R_{IB}\approx 10^3$. 
{At small $R_{IB}$,}
the larger vertical integral scale (with respect to its horizontal counterpart), indicative of a lesser anisotropy for strong rotation and stratification (blue triangles), can be attributed to initial conditions that are isotropic together with, in that range, weak nonlinear coupling. 
Note that, at a given $R_{IB}$, vertical scales are almost a factor of 2 larger for stronger rotation, with a clear clustering of points with $Ro\le 0.3$ (blue triangles) at intermediate values of $R_{IB}$. This can be associated with a  stronger inverse energy transfer due to rotation, although an inverse energy cascade is not directly observed in the absence of forcing, but can appear, for long times, as an envelope to the temporal decay behavior of a turbulent flow\cite{mininni_13}. 

In Fig. \ref{f:3}(b)-(c) are shown the second invariants,  $d_{II}^{1/2}$ and 
$v_{II}^{1/2}$ of the velocity gradient and vorticity tensors (see equation (\ref{tensors}) 
for definitions), again as functions of $R_{IB}$. While anisotropy expressed in terms of $d_{II}^{1/2}$
seems to show an approximate power law decrease towards isotropy (with power law index -1/3), in 
$v_{II}^{1/2}$ the three regimes of mixing are again visible. In the latter, a sharp transition is observed at 
$R_{IB} \gtrsim 100$. 
In terms of Froude number, the intermediate regime is bounded by 
$Fr \in [0.03, 0.2]$, and in terms of $R_B$ it is bounded by $R_B \in [10,300]$. Note that the $Fr$ bounds encompass that for which the 
{large-scale}
intermittency is strongest in the case of purely stratified forced flows, as 
measured by the kurtosis of the vertical Lagrangian velocity\cite{feraco_18} 
(see also Fig. \ref{f:4}). 
Note also that, for the highest values of the 
{buoyancy}
interaction parameter, we have a small $d_{II}^{1/2}\approx 10^{-3}$, whereas in terms of the vorticity tensor, the tendency toward isotropy is much slower, with a lowest value of order $10^{-1}$, indicative of vorticity structures that retain a signature of the imposed anisotropy.

{
This variable anisotropy associated with strong mixing properties is also accompanied by marked intermittency,  at small scales and also in the large scales. We first analyze in Fig. \ref{f:4} the PDFs of the vertical temperature gradients, either normalized (a, left), or without normalization (b, middle). Both are binned in $N/f=Ro/Fr$, and at left the dotted line represents the corresponding Gaussian distribution. As expected,  the PDFs are non-Gaussian for all parameters, with wings the intensity of which varies somewhat with $N/f$. We note however that such wings are present from purely stratified flows to strongly rotating (and stratified) flows with $Ro\le 0.3$. Gradients favor small scales, but large scales are intermittent as well, as found already for purely stratified flows\cite{rorai_14}, at least for an interval of parameters\cite{feraco_18}.  As an example of such large-scale intermittency, we plot at right (c) the kurtosis of the vertical component of the Eulerian velocity, $w$,  at the peak of dissipation, with the kurtosis defined as $K_w=\left<w^4\right>/\left<w^2\right>^2$. When considering only the runs with isotropic initial conditions, the increase in $K_w$ is rather smooth and with a peak at $R_{IB} \approx \mathcal{O}(10)$ of $K_w\lesssim 4$. 
What is particularly striking,
however, is the ``bursty'' behavior seen in the runs with QG initial conditions (indicated by stars)
with a peak of $K_w \approx 7.5$ at $R_{IB} \gtrsim 1$, or at $Fr \approx 0.07$, in 
good agreement with what is found  for forced flows\cite{feraco_18}. The high values we 
see in $K_w$ are comparable to those observed in 
the atmosphere\cite{mahrt_94, lenschow_12}.
{Note however}
that the peaks in $K_w, K_\theta$ are intermittent in time\cite{feraco_18}, whereas our analysis is done at a fixed time close to the maximum dissipation of the flows, in order to maximize the effective Reynolds number of each run.
\\
 The behavior of the QG runs with significantly higher kurtosis is probably due to the fact that their initial conditions are two-dimensional, and with $w=0$. In such  a case, for small Froude number and at least for small times,  the advection term leads to smooth fields, and the flow  has to develop strong vertical excitation 
 characteristic of stratified turbulence, through local instabilities, in order to catch up with energy dissipation and with emerging tendencies towards isotropy in the small scales.
The temperature (not shown) displays for most runs a relatively flat kurtosis at close to 
its Gaussian value, $K_\theta^{(G)} \approx 3$, but still exhibits a rather sharp increase to 
$K_\theta \gtrsim 4.2$ in the QG-initialized runs at $R_{IB} \approx 1$, as well as for smaller values of Froude number and buoyancy interaction parameter
$R_{IB}$.
}

We provide in Fig. \ref{f:5} the parametric variations for some of the  
velocity- and temperature-related anisotropy tensor invariants defined in equ. (\ref{tensors}). Fig. \ref{f:5}(a)-(b) show $b_{II}$ as a function of $R_{IB}$, and $g_{II}$ as a function of  $Ri$, 
respectively. Both have a peak at $R_{IB}\approx 1$,  $Ri \approx 1$ 
(corresponding also to $Fr\approx 0.075$, $R_B\approx 10$, not shown); however, we 
note that $g_{II,III}$ have a maxima for slightly smaller values of $Fr$.
The final transition to a plateau approaching isotropic values, seen in 
Fig. \ref{f:5}(a), occurs for high 
$R_{IB}$ $\approx 10^3$, as advocated on the basis of oceanic and estuary 
measurements\cite{gargett_84}, or from  DNS\cite{smyth_00b, lang_19}.

 \begin{figure*}    
\includegraphics[width=5.9cm,height=6.0cm]{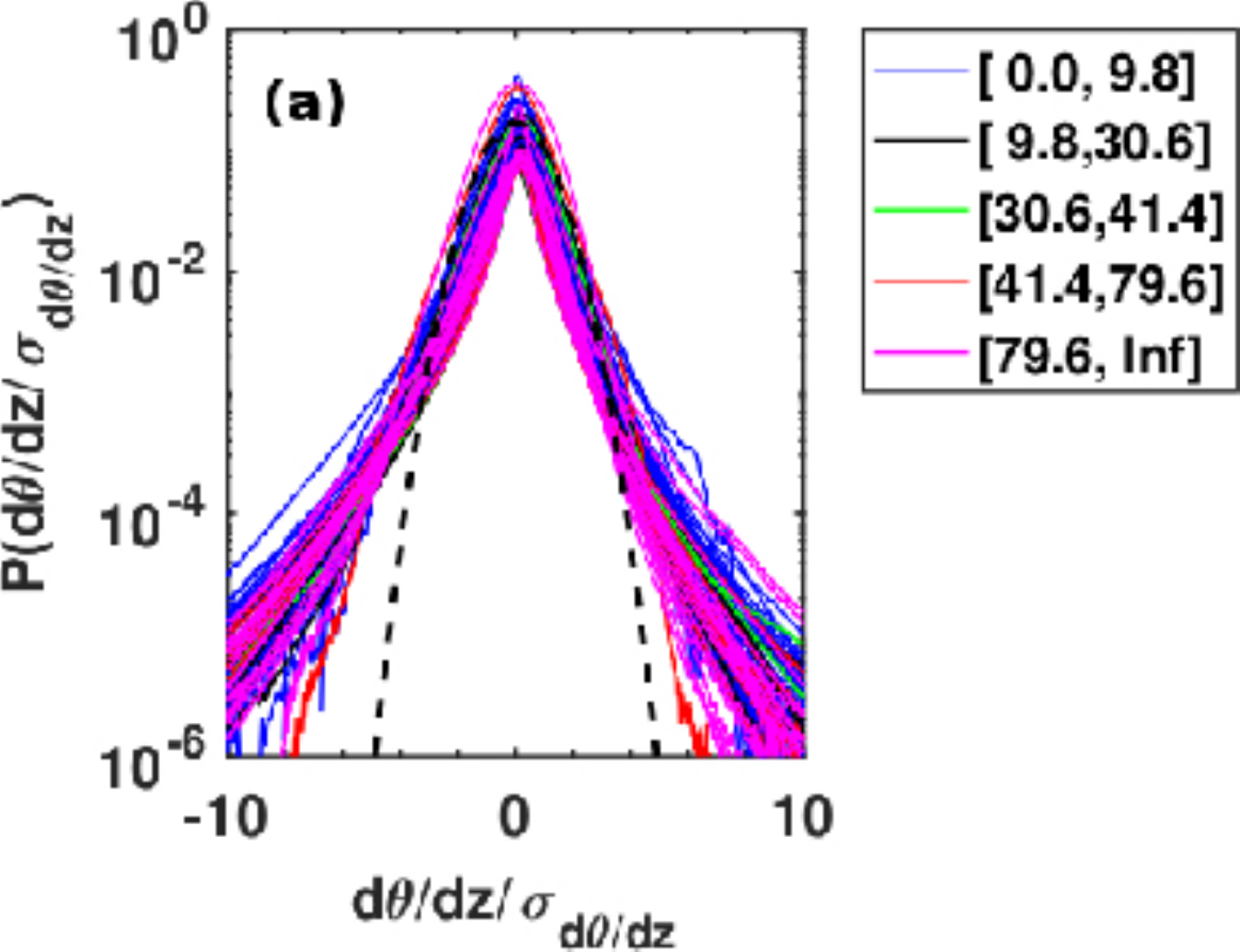}  
\includegraphics[width=5.9cm,height=6.0cm]{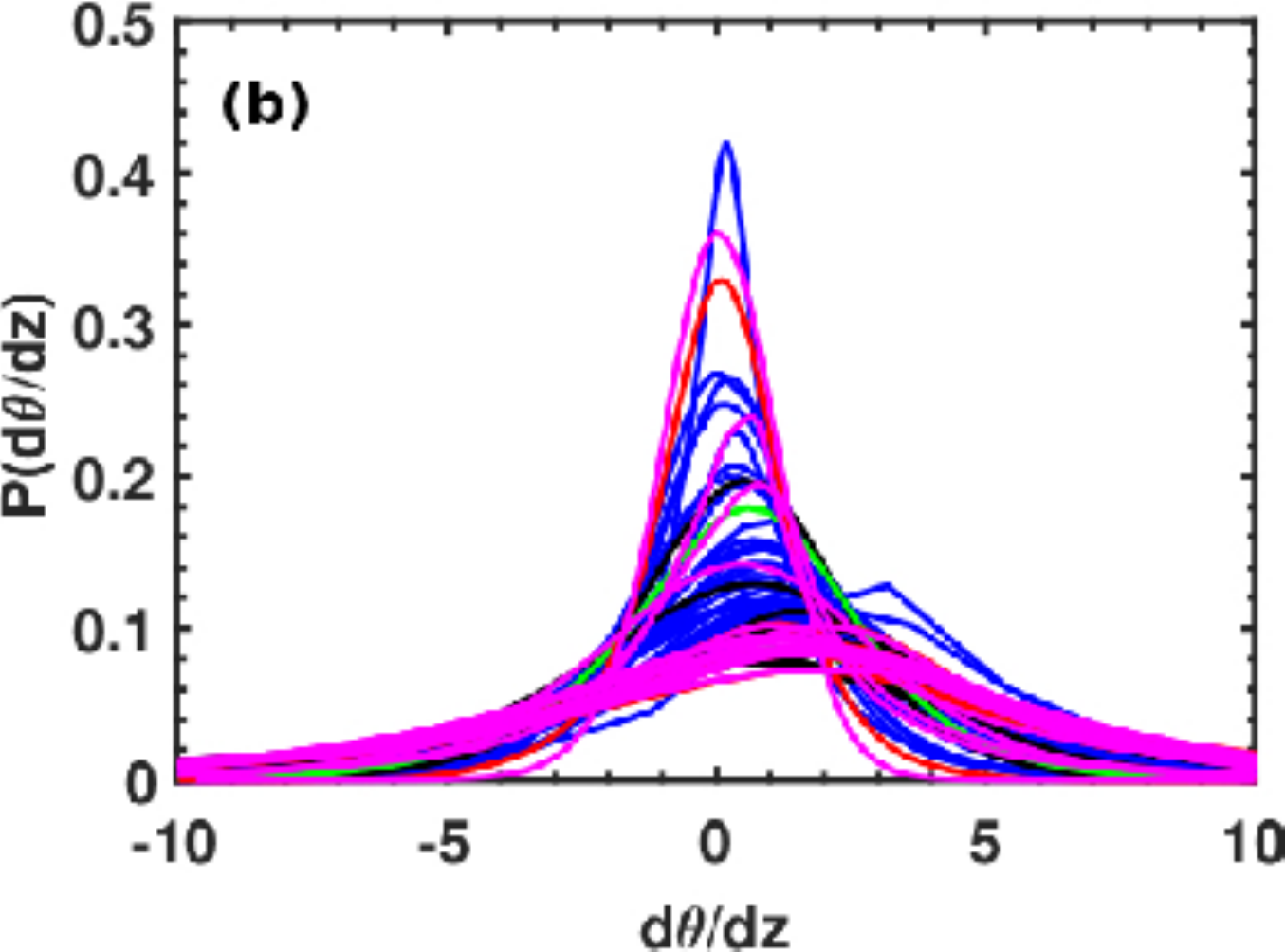}\hskip0.04truein
\includegraphics[width=5.9cm,height=6.0cm]{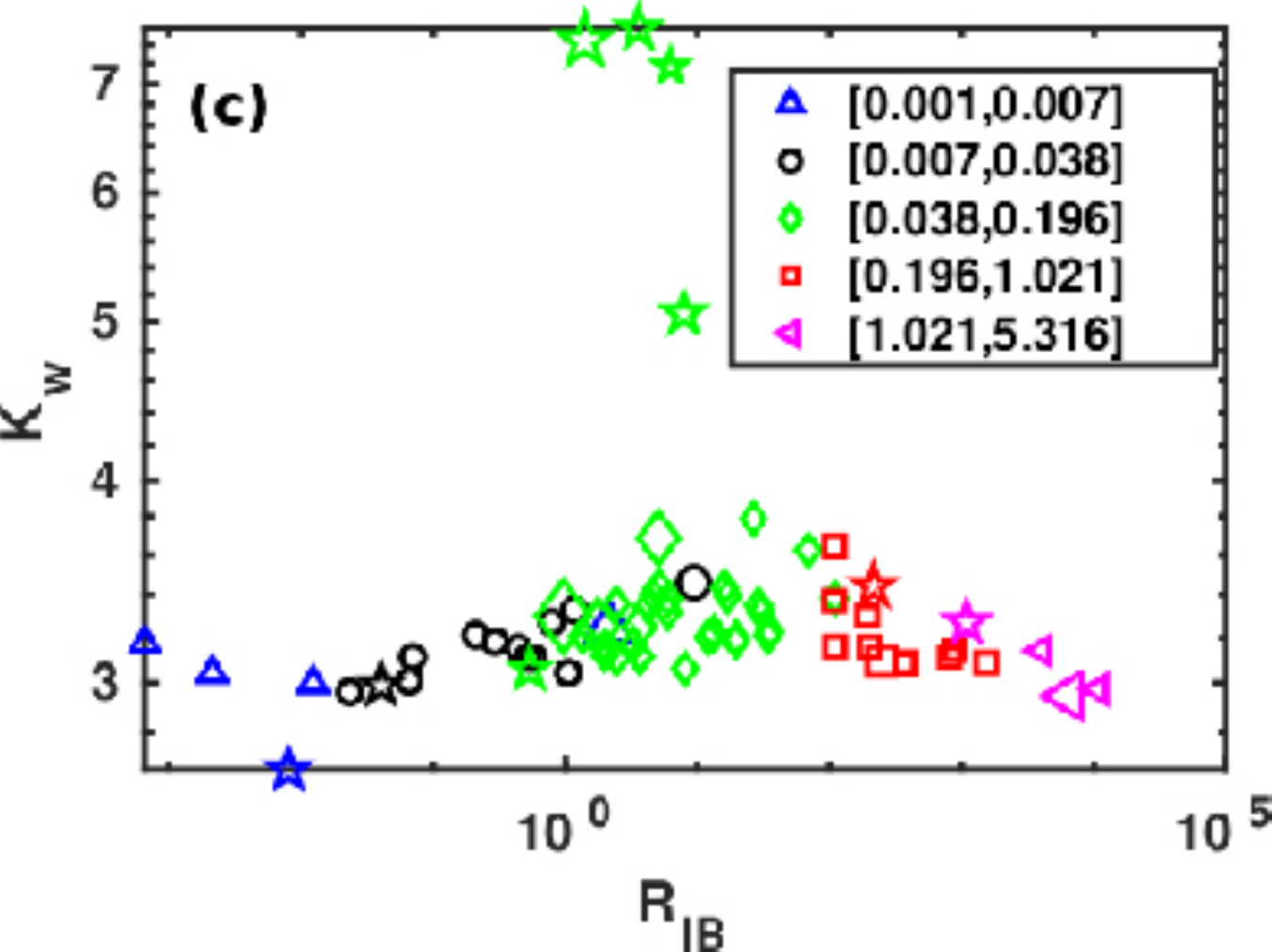}   
\caption{(a): 
{Normalized} 
PDFs of $\partial_z \theta$ with binning in $N/f$ (see legend).
{The dotted black line is the corresponding Gaussian distribution.
(b): The same PDFs without the normalization.}
(c): Kurtosis of vertical velocity as a function of $R_{IB}$, with binning in Froude number as indicated in the legend.
}\label{f:4}    \end{figure*}

Having scaled nonlinearly both the second and third invariants of tensors  in order for them to
have the same physical dimension, we find that third invariants have similar scaling 
with control parameters, except that they can and do become negative, in 
ways comparable to what is found for purely stratified flows\cite{smyth_00b}. We illustrate this
in Fig. \ref{f:5}(c) in a scatter plot of the second and third invariants of
$b_{ij}$ that, to a large degree, fills in Fig. 6 of Ref. \citet{smyth_00b}
for $b_{II}^{1/2} < 0.2$; it highlights the fact that at the peak of enstrophy,
the majority of our runs are dominated by oblate axisymmetric structures,
in the form of sheets. This is complementary to what is performed by using many temporal snapshots, when one can probe more of the permissible $b_{II} - b_{III}$ domain\cite{smyth_00b}.

\begin{figure*} 
\includegraphics[width=8cm]{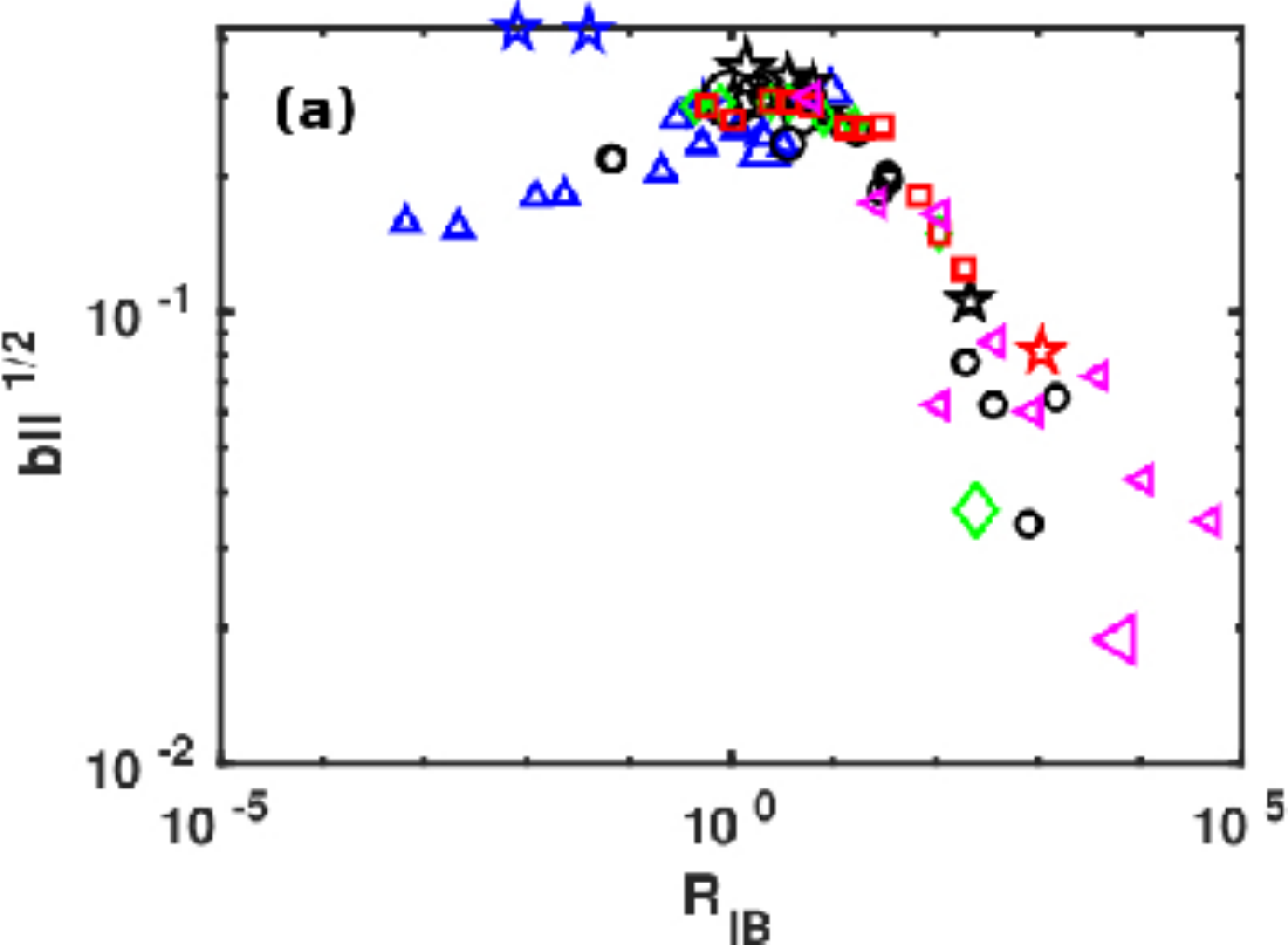} \hskip0.05truein 
\includegraphics[width=8cm]{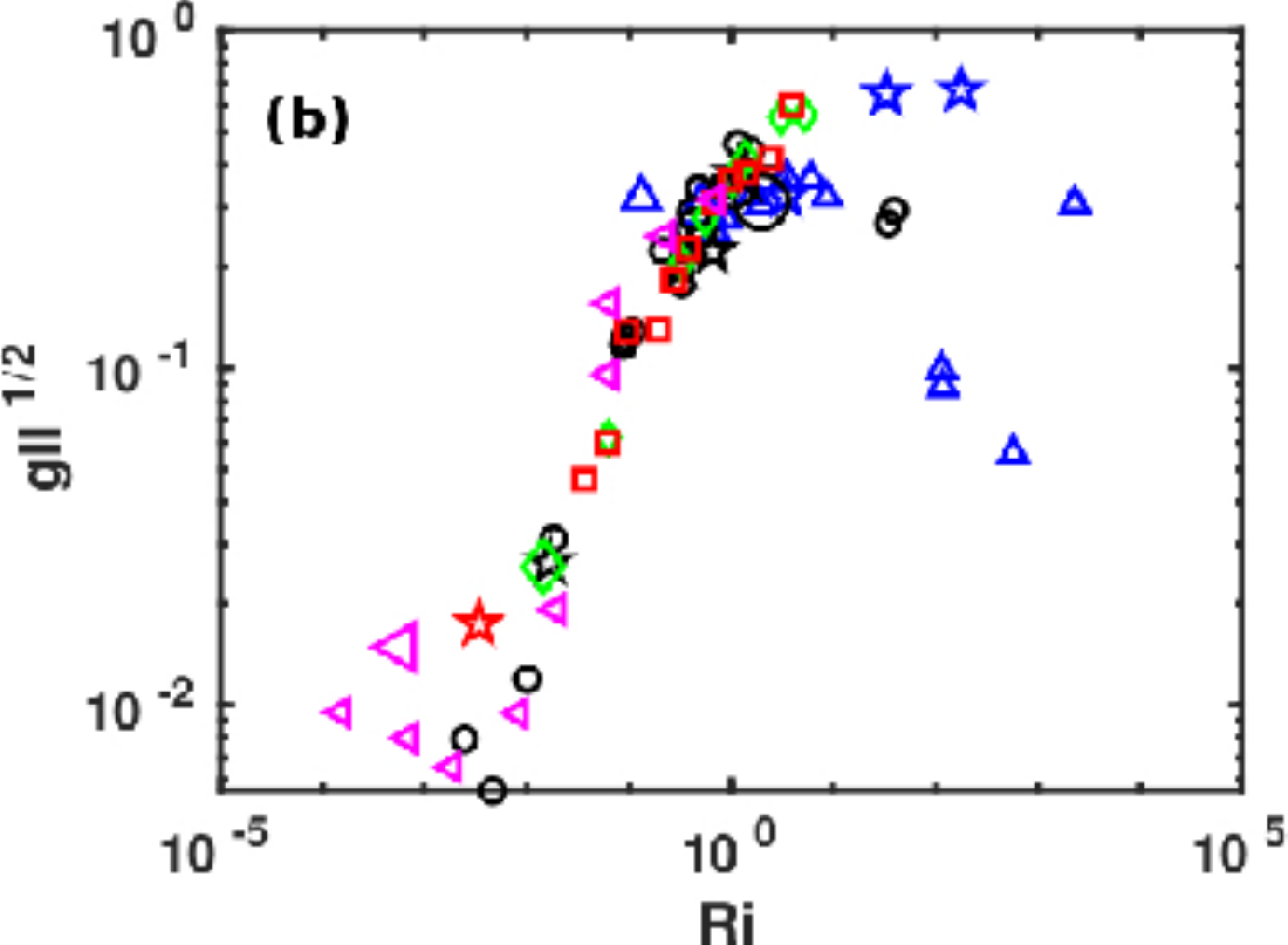} 
\includegraphics[width=8cm]{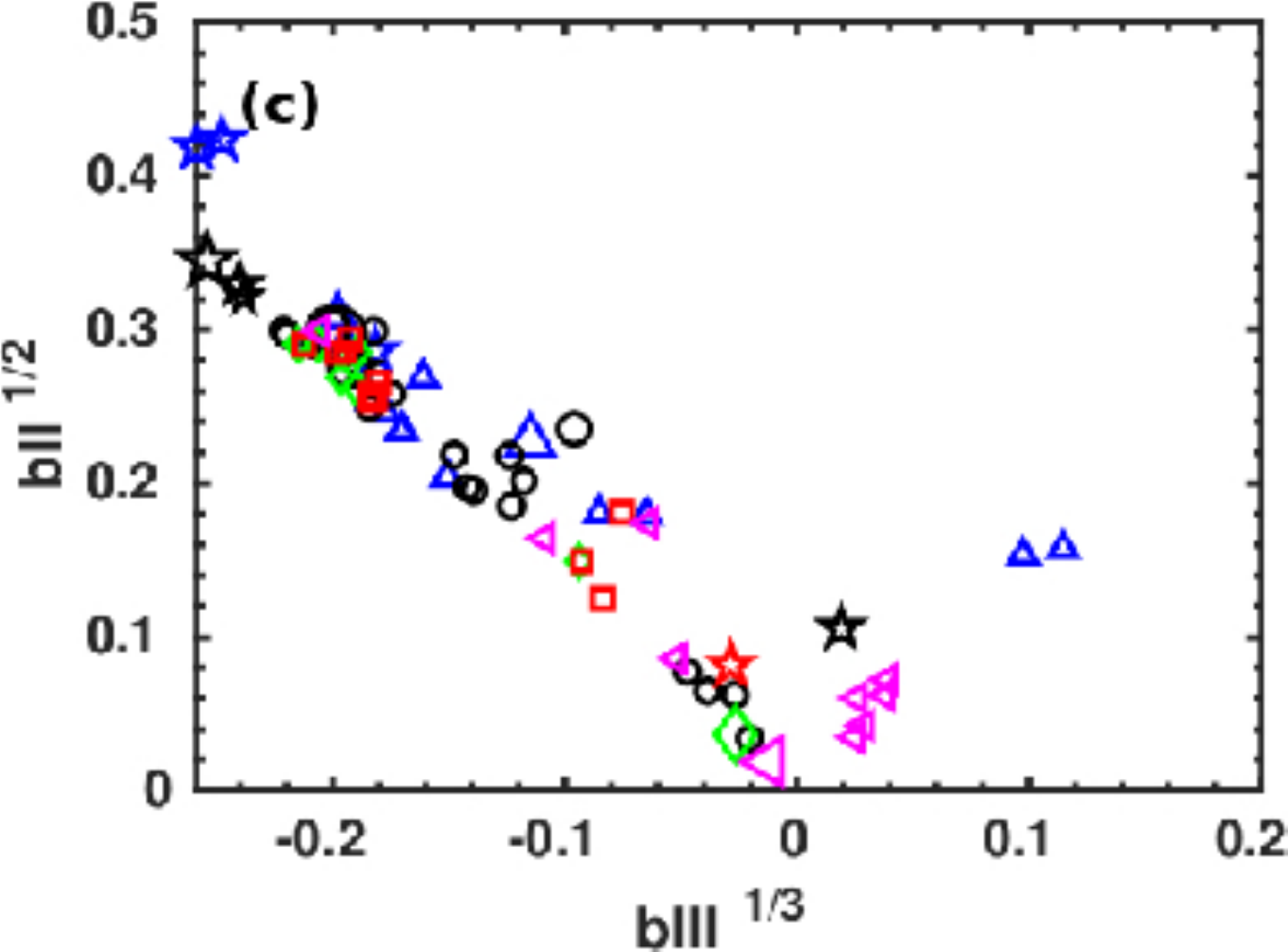}\hskip0.05truein 
\includegraphics[width=8cm]{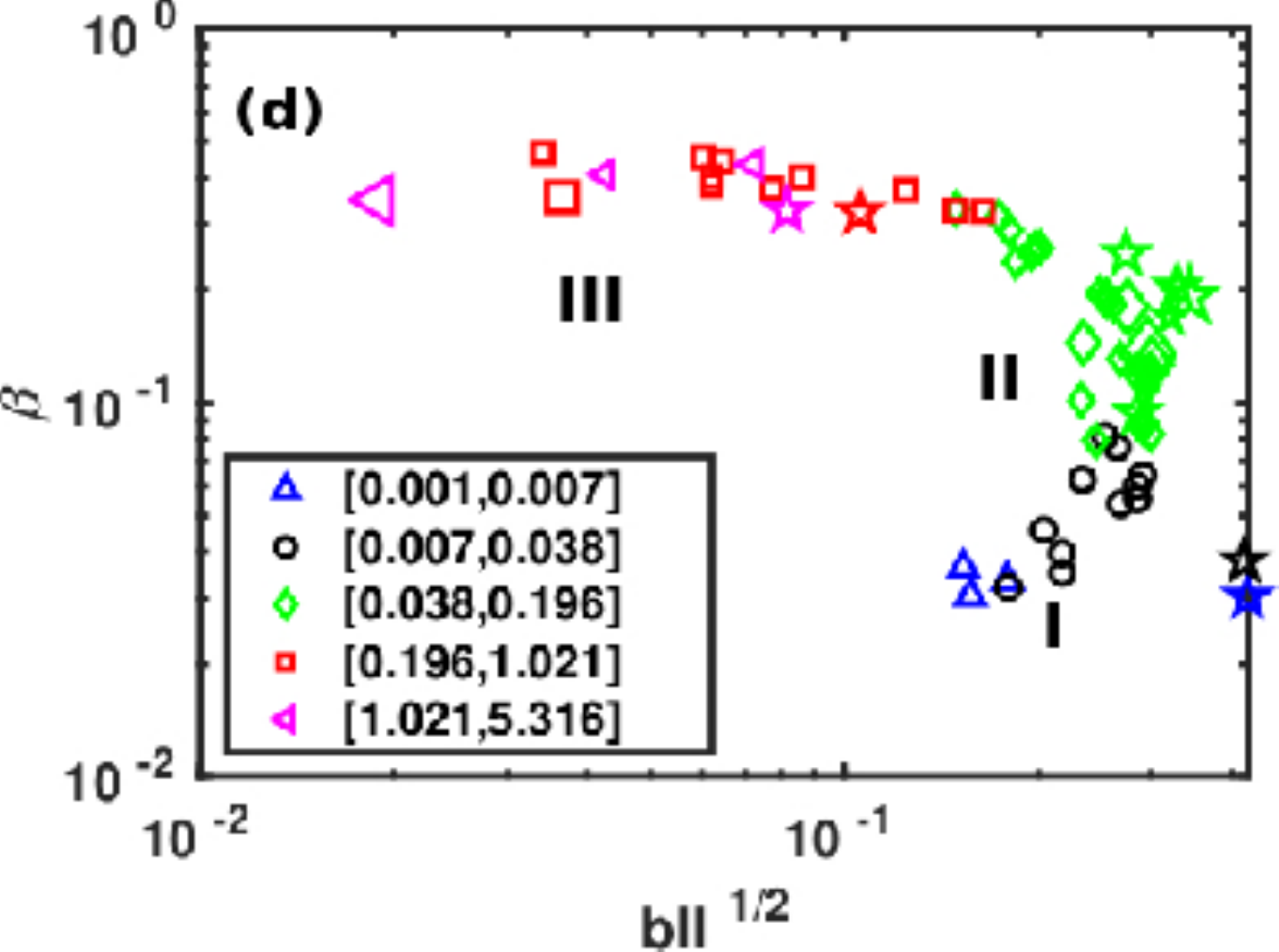} 
\includegraphics[width=8cm]{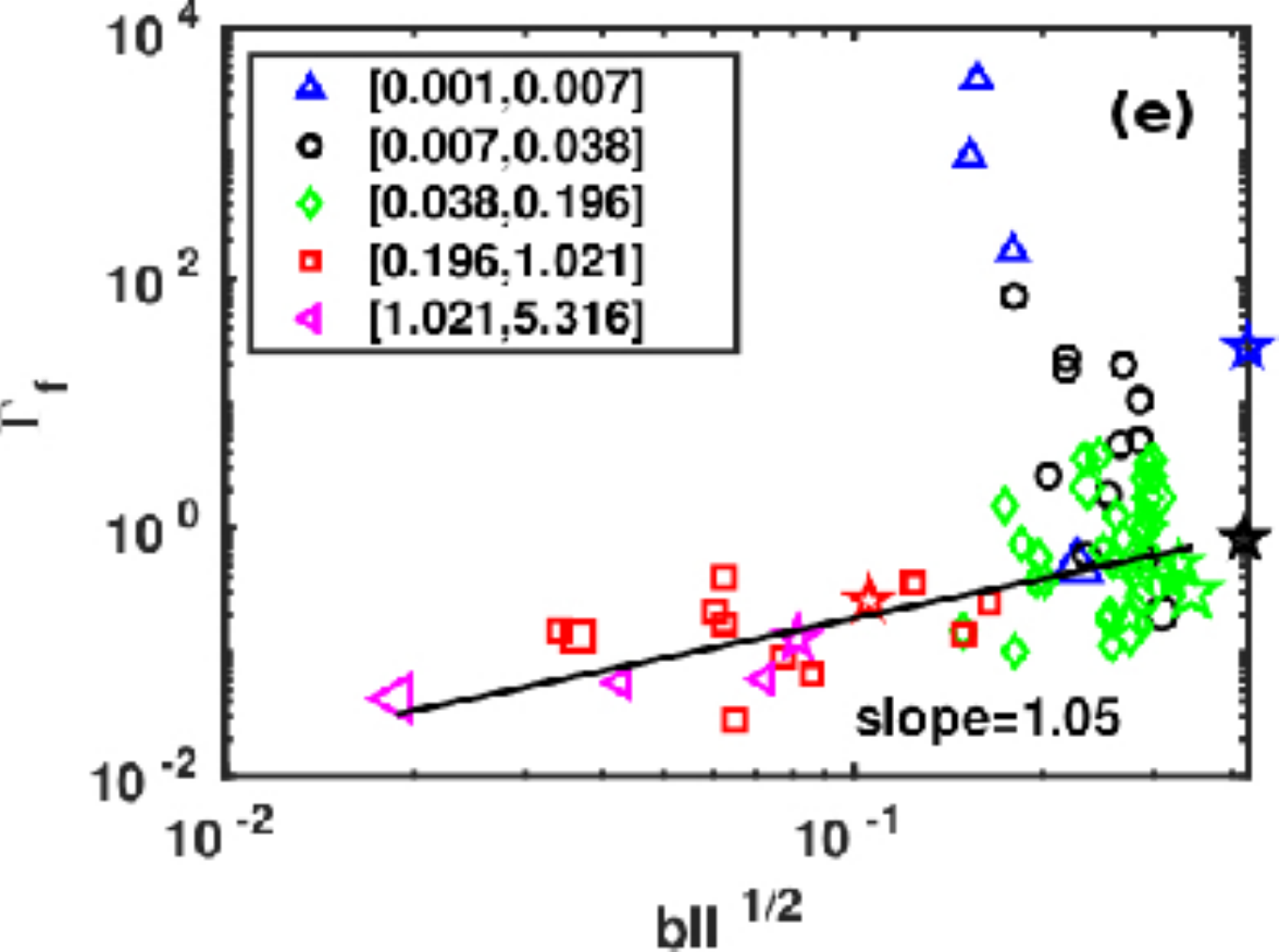} 
\caption{ 
Velocity and temperature invariants defined in eq. (\ref{tensors}):
(a) $b_{II}^{1/2}$ {\it versus} $R_{IB}$; 
(b) $g_{II}^{1/2}$ {\it vs.} $Ri$; 
(c) $b_{II}^{1/2}$ {\it vs.} $b_{III}^{1/3}$; and 
(d) $\beta$ {\it vs.} $b_{II}^{1/2}$, showing the three regimes as
in Fig. \ref{f:1}. 
{In (e) is given the}
mixing efficiency $\Gamma_f$ {\it vs.} $b_{II}^{1/2}$, with a best-fit 
reference line  for $Fr > 0.05$.
In (a,b,c), color binning is done in terms of Rossby number (see inset in Fig. \ref{f:1}), whereas in (d,e) it is in terms of $Fr$.
}\label{f:5}    \end{figure*}

We do note that there are two straggler points at high $b_{II}, g_{II}$, and low negative $b_{III}$, and in
$v_{II}$ as seen in Fig. \ref{f:3}(c). These runs,   indicated by blue stars, have quasi-geostrophic initial conditions and  are at low Froude, Rossby  and buoyancy Reynolds number ($R_B\lesssim 1$); specifically, they are\cite{pouquet_18} runs Q9 and Q10 (see their Table 2). 
Again, the quasi two-dimensional nature of such flows at the peak of enstrophy is confirmed in Fig. \ref{f:5}(c), which places these QG-initialized runs on the upper left branch. 
This indicates that these flows are dominated by quasi two-dimensional sheets\cite{smyth_00b} (see, {\it e.g.}, their Fig. 6). Indeed, the high anisotropy observed in the vicinity of $R_B \approx 1, R_{IB}\approx 1, Fr\approx 0.07$ in Fig. \ref{f:3}(c)  corresponds to two-dimensional structures in the form of shear layers with strong quasi-vertical gradients at low $Fr$, and which eventually roll-up as they become unstable. 

Fig. \ref{f:5}(d) shows the dependence of the kinetic
energy dissipation efficiency, $\beta$, on the second invariant of the velocity 
anisotropy tensor, this time with binning in $Fr$. The figure serves to 
complement both Fig. 1(c) in Ref. \citet{pouquet_18} and Fig. \ref{f:1} (right), 
illustrating behavior in the three RST regimes, where $\beta$ is low at reasonably 
high measures of (large-scale) anisotropy (as measured by $b_{ij}$) in regime I, 
approaches its highest value at largely constant $b_{II}^{1/2}$ in regime II, 
and as anisotropy begins to diminish at the end of regime II, remains 
essentially constant in regime III, as the anisotropy continues to decrease
with decreasing stratification.

Finally, in order to render more explicit the correlation between mixing and anisotropy, we
show in Fig. \ref{f:5}(e) the mixing efficiency $\Gamma_f$, displayed against 
$b_{II}^{1/2}$, again
with binning in Froude number.  One observes an approximate power law increase 
in mixing efficiency as anisotropy grows with stratification, from large to 
moderate $Fr$, with a best fit slope of $\approx 1$. Using the definitions for
$\beta$ and $\Gamma_f$ in terms of the buoyancy flux, we can write
$\Gamma_f =[\beta Fr]^{-1} \left< w \theta \right>/\left<u^2\right>$. Noting again that in regime III,
$\beta$ is independent of anisotropy (Fig. \ref{f:5}(d)), and that 
$b_{zz}$ is remarkably linear in (indeed, nearly equal to)  
$b_{II}^{1/2}$ for all runs (not shown),  the power law dependence of 
$\Gamma_f$ on $b_{II}^{1/2}$ mainly results
from the increasingly passive nature of the scalar in transitioning from 
regime II to 
{regime}
III, and continuing to larger $Fr$.
There is also an abrupt increase in $\Gamma_f$ in the smallest $Fr$ range, corresponding to regime I with negligible kinetic energy dissipation. 
The transitory regime (green diamonds) in the vicinity of the peak of vertical velocity kurtosis also corresponds to maximum $b_{II}$, {\it i.e.} maximum anisotropy, together with mixing efficiency of order unity. 
The accumulation of points for Froude numbers in the intermediate range of values has large $b_{II}$ and a mixing efficiency around unity, with quasi-balanced vertical buoyancy flux and kinetic energy dissipation.


 \section{Conclusion and discussion}  
  
We have shown in this paper   that, in rotating stratified turbulence, a sharp increase in dissipation and mixing efficiency  is associated, in an intermediate regime of parameters, with large-scale anisotropy,
{as seen in the velocity tensor $b_{ij}$,}
 and large-scale intermittency,
 {as observed in the vertical velocity through its kurtosis.}
  The  return to isotropy is slow and  takes place mostly for buoyancy 
{interaction} 
parameters 
{$R_{IB}$} 
larger than $\approx 10^3$.
{The temporal persistence of anisotropic structures at a given Froude number could be related to the slow decay of energy in the presence of waves, particularly when helicity is strong\cite{rorai_13}. Furthermore, }
rotation plays a role in the large scales, with a larger vertical integral scale at a given Froude number for small Rossby numbers (see Fig. \ref{f:3}(a)). The return to large-scale isotropy, as measured by ${L_z}/{L_\perp}\approx 1$, is very sharp. 
These results evoke threshold behavior and avalanche dynamics, as analyzed for numerous  physical systems\cite{chapman_01, uritsky_10, uritsky_17, sorriso_19} in the context of  the solar wind, and as found as well recently in observational oceanic data\cite{smyth_19}.

 In order to determine whether a given system is undergoing self-organized criticality (SOC) in the form of so-called avalanches, and if so what SOC class the system belongs to, one needs to resort to spatio-temporal analysis, although proxies are possible, 
{such as using static snapshots of dissipative structures  and applying the law of probability 
conservation\cite{uritsky_10}.}
 Furthermore, different conclusions may be drawn whether one examines structures in the inertial range of  turbulent flows, or whether one is in the dissipative range\cite{uritsky_10}. Perhaps localized Kelvin-Helmoltz overturning vortices merge into larger regions, as a reflection of non-locality of interactions in these flows, together with  sweeping of small eddies by large-scale ones, close to the linear instability for $Ri_g=1/4$, and leading to rare large-amplitude dissipative (avalanche) events. In that context, long-time dynamics, in the presence of forcing, should be investigated to see whether correlations emerge. A threshold analysis could be performed  in these flows in terms of the number of excited sites, say above a local dissipation rate ${\bar{\epsilon}}_C$, as a function of a control parameter, likely the local gradient Richardson number. Temporal dynamics should also be analyzed in terms of life-time of overturning structures, as performed classically for example for pipe flows\cite{wangj_15, lemoult_16}.

The burstiness of these rotating stratified flows is accompanied by  a turbulence collapse.
{This takes place} 
once the energy has been dissipated at a rate close to that of homogeneous isotropic turbulence.
However, we note that this rate has been found to be  dependent on the ratio of the wave period controlling the waves to the turn-over time  
{(in other words, the Froude number),}
in an intermediate regime of parameters\cite{pouquet_18}.  
This type of behavior has been studied {\it e.g.} for shear flows, emphasizing both the inter-scale interactions between large and small eddies with rather similar statistics\cite{shih_16}, 
{and}
 the importance of sharp edges in frontal dynamics\cite{pomeau_86}. 
This has been analyzed  in the laboratory at the onset of instabilities including for Taylor-Couette flows or for pipe flows \cite{lemoult_16, dessup_18},  and it may be related to frontal dynamics observed in the atmosphere and ocean\cite{mcwilliams_16, sujovolsky_18}, given the tendencies of such flows to be, at least in the idealized dynamical setting studied herein, at the margin of such instabilities.

Recent observations\cite{smyth_19}, modeling\cite{mcwilliams_16} and numerous DNS indicate that indeed the gradient Richardson number resides mainly around its classical threshold for linear instability ($\approx 1/4$), as  also observed in our results, exhibiting a strong correlation with dissipation. 
In that light, it may be noted that the range of parameters for the mixing efficiency to be comparable to its canonical value observed in oceanic data is close to the instability threshold: $\Gamma_f\approx 0.2$ for $0.02 \lesssim Fr \lesssim 0.1$. Similarly, the kurtosis of the temperature and vertical velocity $K_{\theta,w}$  are high\cite{feraco_18} in a narrow window around $0.07 \le Fr \le 0.1$.
As a specific example of marginal instability behavior in the framework of a classical model of 
turbulence\cite{vieillefosse_82, meneveau_11} extended to the stratified case, it can be shown\cite{sujovolsky_19} that the flow remains close to the stable manifold of a reduced system of equations governing the temporal evolution of specific field gradients, involving in particular the vertically sheared horizontal flows through the second and third invariants of the velocity gradient matrix, and a cross-correlation velocity-temperature gradient tensor. 

The link between 
{localized}
 intermittency, anisotropy and dissipation  is also found in fully developed turbulence, in the form of strong vortex filaments, non-Gaussianity of velocity gradients and localized dissipative events. 
 {It has also been shown that a Kolmogorov spectrum $E_v(k)\sim k^{-5/3}$ can still be observed when the small scales are anisotropic\cite{elsinga_16}.}
 The new element in rotating stratified flows is what the wave dynamics brings about, namely a fluid in a state of marginal  instability, almost everywhere close to the threshold of linear instability in terms of $Ri_g \approx 1/4$.
It is already known that in magnetohydrodynamics (MHD), when coupling the velocity to a magnetic field leading to the propagation of Alfv\'en waves, there is stronger 
{small-scale}
intermittency than for FDT, as found in models of MHD\cite{grauer_94, politano_95a}, 
in DNS\cite{politano_98b, mininni_06b} as well as in observations of the solar wind\cite{marino_12}. In RST, the added feature is 
having intermittency in the vertical component of the velocity and temperature fluctuations themselves, thus at large scale, as found in many observations in the atmosphere and in climatology as well\cite{petoukhov_08, sardeshmukh_15}, and limited to a narrow range of parameters\cite{feraco_18} centered on the marginal instability threshold. 
Thus, not only does this interplay between waves and nonlinear eddies not destroy these characteristic features of turbulent flows, but in fact it acts in concert with them and can rather enhance them as well.

The large data base we use is at a relatively constant Reynolds number, $Re\approx 10^4$, and thus an analysis of the variation of anisotropy with $Re$, for fixed rotation and stratification remains to be done, in the spirit of earlier pioneering studies\cite{cambon_89, antonia_94} for fluids. Also, scale by scale anisotropy might be best studied with Fourier spectra. This will be accomplished in the future, together with a study of the role of forcing. 

This paper is centered on a large parametric study of rotating stratified turbulence. Each flow taken individually is strongly intermittent in space, and thus presents zones that are active as well as zones that are
quiescent. It was proposed recently to partition a given flow in such zones, with strong layers delimiting such patches\cite{portwood_16}, depending on the buoyancy interaction parameter $R_{IB}$, and with threshold values,
 of roughly 1, 10 and 100. The intermediate range corresponds, in our DNS runs, to the peak of anisotropy and intermittency together with mixing efficiency being close to its canonical value, $\Gamma_f\approx 0.2$. In 
 {this regard,}
  it will be of interest to perform such a local study for a few given runs of our data base in the three regimes.

Many other extensions of this work can be envisaged. For example, one could  perform  a wavelet decomposition to examine the scale-by scale anisotropy  and intermittency in such flows\cite{bos_07}.
Moreover, kinetic helicity, the correlation between velocity and vorticity, is created by turbulence in rotating stratified flows\cite{hide_76, marino_13h}. It is the first breaker of 
{isotropy,}
 since flow statistics depend only on the modulus of wavenumbers, but two defining functions (energy and helicity density) are necessary to fully describe the dynamics. In FDT,  helicity is slaved to the energy in the sense that $H_v(k)/E_v(k)\sim 1/k$, {\it i.e.} isotropy is recovered in the small scales at the rate $1/k$.
In the stratified case, its scale distribution  changes with \BV\cite{rorai_13}, as measured for example in the PBL \cite{koprov_05}, and  it undergoes a direct cascade  to  small scales while  energy goes to  large scales in the presence of strong rotation and forcing\cite{mininni_09c}. What role helicity and the nonlinear part of potential vorticity, namely 
$\vomega \cdot \nabla \theta$, will play in the fast destabilization of shear layers, their intermittency, anisotropy and criticality are topics for future work.  

We conclude by noting that a deeper understanding of the structure of small-scale rotating stratified turbulence, 
and of the nonlocal interactions between small scales and large scales,
will allow for better modeling in weather and climate codes. Many models of anisotropic flows have been 
proposed\cite{mahrt_14, giddey_18, maulik_18, heinz_19} 
{including artificial neural networks. They extend }
isotropic formulations for kinetic energy dissipation by adding several off-diagonal terms, and assuming (or not) isotropy in the orthogonal plane\cite{thoroddsen_92, werne_01},  
{including}
for two-point closures\cite{sagaut_08a}. 
{This modeling strategy}
has already been found useful in models of  turbulent mixing in oceanic simulations\cite{wijesekera_03, belcher_12}.

\begin{acknowledgments}
The runs analyzed in this paper have been performed using an ASD allocation at NCAR, supplemented by a large amount of background time, for both of which  we  are thankful.
NCAR  is funded by the National Science Foundation.
RM acknowledges support from the program PALSE (Programme Avenir Lyon Saint-Etienne) of the University of Lyon, in the framework of the program Investissements d'Avenir (ANR-11-IDEX-0007), and from the project "DisET" (ANR-17-CE30-0003).
Support for AP from LASP, and in particular from Bob Ergun, is gratefully acknowledged.
\end{acknowledgments}  

\section{Appendix}
\label{S:appendix}

We put together in the following Table, for convenience, the various symbols which are used in the paper, with their definitions and names. 

\begin{table*} \begin{ruledtabular}    \begin{tabular}{lccccccccccc}
{Symbol} & {Definition} & {Description} &  {Remarks} & \\ 
\hline  \hline
${\bf u}$& & Velocity fields of components $[u_\perp, w]$ \\
$\vomega$& $\vomega=\nabla \times {\bf u}$ & Vorticity field \\
$S$ & $\left< \partial_zu_\perp \right>$ & Internal mean vertical shear  &  Of local density $S({\bf x})$ \\
$\theta$  &  & Temperature fluctuations \\
$P$ &  & Pressure \\
${\bf k}$ & & Wavevector of modulus $k= [{\bf k}_\perp^2  +  k_z^2]^{1/2} $ \\
\hline
$U_0$  &  & Characteristic velocity  \\
$E_v(k)$  & $\int E_v(k)dk=\left<|{\bf u}|^2/2 \right>$ & Isotropic kinetic energy spectrum \\ 
$L_{int}$  & $2\pi \int [E_v(k)/k]dk/E_v$  & Integral scale  \\
$E_p(k)$  & $\int E_p(k)dk=\left<\theta^2/2 \right>$ & Isotropic potential energy spectrum  &   Total energy $E_T=E_v+E_p$ \\ 
$N$  &  & \BV  \\
$f$ & $2\Omega$  &  Rotation frequency & $\Omega$: imposed rotation rate  \\
$\nu$  &  & Viscosity  \\
$\kappa$  &  & Diffusivity  \\
\hline
$Z_v$ & $\left< |\vomega|^2 \right>$ &Kinetic enstrophy  \\
$Z_p$ & $ \left< |\nabla \theta |^2 \right>$ &Potential enstrophy &  Total enstrophy $Z_T=Z_v+Z_p$  \\
${\bar {\epsilon}}_v$  &$\nu Z_v$ &  Mean kinetic energy dissipation \\
${\bar {\epsilon_p}}$ & $\kappa Z_p$ & Mean potential energy dissipation    & Total dissipation ${\bar {\epsilon}}_T={\bar {\epsilon}}_v + {\bar {\epsilon}}_p$ \\
$\epsilon_D$ & $U_0^3/L_{int}$ & Dimensional kinetic energy dissipation \\
$s_{ij}({\bf x})$ & $(\partial_i u_j + \partial_j u_i)/2$ & Point-wise strain rate tensor  & $\epsilon_v ({\bf x})= 2\nu s_{ij}({\bf x}) s^{ij}({\bf x}) $ \\
\hline
$\eta$ & $[\nu^3/\epsilon_v]^{1/4}$ & Kolmogorov (dissipation) scale \\
$\tau_{NL}$  & $L_{int}/U_0$ & Dimensional eddy turn-over time     \\
$T_v$  & $E_v/\epsilon_v$ & Effective transfer and dissipation time  & for the kinetic energy     \\
$T_p$  & $E_p/\epsilon_p$ & Effective transfer and dissipation time   & for the potential energy    \\
\hline \hline
Re & $U_0L_{int}/\nu$ & Reynolds number  &  {\bf Four Governing Parameters} \\
Fr & $U_0/(L_{int} N)$ & Froude number \\
$Ro$ & $U_0/(L_{int}f)$ &  Rossby number \\ 
Pr & $\nu/\kappa$ & Prandtl number &  {\sl Equal to unity for all runs} \\
\hline 
$R_B$ & $Re Fr^2$ & Buoyancy Reynolds number  & {\bf Four Derived Parameters}\\
$R_{IB}$ & $ {\bar{\epsilon}}_v/(\nu N^2)$ & Buoyancy interaction parameter &  At times called buoyancy parameter\cite{ivey_08} \\ 
$Ri$ & $N^2/S^2$ &  Richardson number  &    \\ 
$Ri_g$ & $N(N-\partial_z\theta)/S({\bf x})^2$ & Point-wise gradient Richardson number & Close to instability threshold \\
\hline  \hline 
$B_f$  & $\left< w \theta \right>$ & Vertical buoyancy flux      \\
$\Gamma_f$  & $B_f/{\bar {\epsilon}}_v$ & Mixing efficiency      \\
$\Gamma_\ast$  & ${\bar {\epsilon_p}}/{\bar {\epsilon_v}}$ & Reduced mixing efficiency \\
$\ell_{Oz}$  & $2\pi [{\bar {\epsilon_v}}/(N^3)]^{1/2}$ & Ozmidov scale & $\ell_{Oz}/\eta\sim R_{IB}^{3/4}$   \\

\hline 

$L_{Ell}$  & $2\pi \theta_{rms}/N$ & Ellison scale  &  $L_{Ell}/L_{int}\sim Fr$ in regime II  \\
$\beta$  & ${\bar {\epsilon}}_v/\epsilon_D$ & Dissipation efficiency  & $\beta\sim Fr$ in regime II  \\
 &  & $\beta$ quantifies the changing role of the waves & See \citet{pouquet_18}  \\  
\hline
%
         
         $b_{ij}$& $\left<u_iu_j\right>/\left<u_ku_k\right> - \delta_{ij}/3$ & Velocity tensor, zero for isotropy  & Invariants $b_{II}$, $b_{III}$   \\
         $d_{ij}$& $\left< \partial_k u_i \partial_k u_j \right>/\left< \partial_k u_m \partial_k u_m \right> - \delta_{ij}/3$ & Velocity gradient tensor, zero for isotropy & Invariants $d_{II}$, $d_{III}$   \\
$v_{ij}$& $ \left<\omega_i \omega_j \right>/\left<\omega_k \omega_k \right> -\delta_{ij}/3$ 
& Vorticity  tensor, zero for isotropy & Invariants $v_{II}$, $v_{III}$   \\
$g_{ij}$& $  \left< \partial_i \theta\  \partial_j \theta \right>/\left< \partial_k \theta\  \partial_k \theta \right>
         -  \delta_{ij}/3$  & Temperature gradient tensor, 0 for isotropy & Invariants $g_{II}$, $g_{III}$   \\ \\
\hline
$K_w$ & $\left<w^4 \right>/\left<w^2 \right>^2 $ &  Kurtosis (of vertical velocity) & 3 for a Gaussian PDF \\
\end{tabular}  \end{ruledtabular}
\caption{ {List of most symbols used in the paper, with definitions and names (refer as well to the main text).
One can also define perpendicular and vertical integral scales, when using the spectra based on perpendicular and vertical wavevectors (see main text). Also note that, in the third regime of rotating stratified turbulence, $R_B=R_{IB}$ because now  ${\bar {\epsilon}}_v=\epsilon_D$.}}
 \label{appendix} \end{table*}

\bibliography{ap_19_july19}    
  \end{document}